\documentclass[12pt, preprint]{aastex}
\usepackage{natbib}
\usepackage{amsmath}
\begin{document}
\title{A Numerical Study of Long-Range Magnetic Impacts during Coronal Mass Ejections}
\author{M. Jin\altaffilmark{1,2}, C. J. Schrijver\altaffilmark{1}, M. C. M. Cheung\altaffilmark{1}, M. L. DeRosa\altaffilmark{1}, N. V. Nitta\altaffilmark{1}, A. M. Title\altaffilmark{1}}

\altaffiltext{1}{Lockheed Martin Solar and Astrophysics Lab, Palo Alto, CA 94304, USA; jinmeng@lmsal.com}
\altaffiltext{2}{NASA/UCAR LWS Jack Eddy Fellow}

\begin{abstract}
With the global view and high-cadence observations from SDO/AIA  and STEREO, many spatially separated solar eruptive events appear to be coupled. However, the mechanisms for ``sympathetic" events are still largely unknown. In this study, we investigate the impact of an erupting flux rope on surrounding solar structures through large-scale magnetic coupling. We build a realistic environment of the solar corona on 2011 February 15 using a global magnetohydrodynamics (MHD) model and initiate coronal mass ejections (CMEs) in active region (AR) 11158 by inserting Gibson-Low analytical flux ropes. We show that a CME's impact on the surrounding structures depends not only on the magnetic strength of these structures and their distance to the source region, but also on the interaction between the CME with the large-scale magnetic field. Within the CME expansion domain where the flux rope field directly interacts with the solar structures, expansion-induced reconnection often modifies the overlying field, thereby increasing the decay index. This effect may provide a primary coupling mechanism underlying the sympathetic eruptions. The magnitude of the impact is found to depend on the orientation of the erupting flux rope, with the largest impacts occurring when the flux rope is favorably oriented for reconnecting with the surrounding regions. Outside the CME expansion domain, the influence of the CME is mainly through field line compression or post-eruption relaxation. Based on our numerical experiments, we discuss a way to quantify the eruption impact, which could be useful for forecasting purposes.
\end{abstract}


\section{Introduction}
The term ``sympathetic solar events" refers to sequences of eruptions from distinct regions on the Sun with apparently causal relations. Sympathetic solar events were reported even before the space age (e.g.,\citealt{richardson51, becker58}). However, due to a lack of observational evidence, the coupling mechanism remained speculative. With the launch of Solar TErrestrial RElations Observatory (STEREO; \citealt{kaiser05}) and Solar Dynamics Observatory (SDO; \citealt{pesnell12}), we have, for the first time, a (nearly) complete coverage of the Sun from three different perspectives\footnote{The complete coverage started in 2011 and will continue if both STEREO spacecraft continue to function and until they drift past the quadrature points once again around mid-May 2019.}, which gives us an unprecedented opportunity to investigate sympathetic events on a global scale. A dozen solar sympathetic events have been reported so far in solar cycle 24, which are typically in the form of coupled quiescent filament eruptions (e.g., \citealt{schrijver11a, yang12}) or coupled active region/filament eruptions (e.g., \citealt{schrijver11b, schrijver13, shen12}). Compared with the extensively studied initiation mechanisms for isolated events (see, e.g., \citealt{forbes06, kliem14}, and references therein), the mechanisms by which sympathetic events are coupled remain largely unknown. Few allegedly sympathetic events have been extensively analyzed/modeled, so the physical mechanisms of how perturbations propagate from one region to another and how they interact with the background magnetic field and trigger an eruption remain unknown. 

Sympathetic events may also present important implications in understanding the space weather, whose two important elements, non-recurrent geomagnetic storms and solar energetic particles (SEPs), are largely attributable to coronal mass ejections (CMEs, see, e.g., \citealt{gosling93}).  When propagating into the solar wind, sympathetic events are prone to CME-CME interaction (e.g., \citealt{lugaz08, lugaz12, liuy14a}).  Such interactions may significantly modify the structure of the CME-driven shock wave and the properties of the interplanetary CME (ICME), and therefore affect the potential of space weather effects (e.g., \citealt{liu12, mostl12, liuy14b}). Interactions between CMEs from the same active region have often been discussed (e.g., \citealt{lugaz05b, lugaz07, xiong06, lugaz13}), but CMEs can be spatially extended and those from distant regions may also interact, adding complexity to solar wind data (e.g., \citealt{temmer12}).  It is also possible that sympathetic events may play a role in SEP events observed at widely separate locations (e.g., \citealt{richardson14, gomez15}).

Statistical studies suggest that coupling between near-simultaneous events likely exists. By investigating all M5 and above flares observed by SDO using superposed-epoch analysis, \citet{schrijver15} found increased rate of flaring and filament eruptions within the first 4 hours, even at locations more than 20$^\circ$ away from the primary flare. \citet{fu15} found an increase in the M- and X-class flaring rates following new active region emergence. A recent study finds that 90\% and 79\% of X-flares occurred in clusters of 2 or more with the mean separation of $\sim$1 day in solar cycle 22 and 23, respectively (A. M. Title, private communication). These studies support the existence of solar sympathy but do not identify the coupling mechanisms. In this study, we would like to establish whether such coupling really exists, and if so, the physical mechanisms responsible.

To uncover the coupling mechanisms in solar sympathetic events, two approaches have typically been used in past studies. One approach involves using a topological analysis of the magnetic field, as applied to coronal field extrapolations that employ photospheric magnetograms as boundary conditions (e.g., PFSS: potential field source surface model). One well-studied example is the series of events occurring on 2010 August 1-2. A sequence of apparently coupled eruptions was observed at widely separated locations that spanned a full hemisphere of the Sun. From a comprehensive analysis of the observational data from SDO and STEREO for the 2010 August 1-2 eruptions, \citet{schrijver11a} argued that the three active regions involved are connected by topological elements present in the magnetic field  (i.e., separatrix surfaces, separators, and quasi-separatrix layers). A detailed topological analysis of the source-surface background field by \citet{titov12} also strongly supports the idea that these structural features were involved in generating sympathetic eruptions for the 2010 August 1-2 events. The other approach is through numerical modeling. \citet{torok11} reproduce some important aspects of the global sympathetic event on 2010 August 1 using an idealized arrangement of flux ropes in a zero-$\beta$ simulation, and found that the presence of a pseudo-streamer is important for producing the ``twin-filament" eruptions seen in the observations.  Also, \citet{lynch13} present a 2.5D magnetohydrodynamics (MHD) simulation of sympathetic magnetic breakout eruptions from a pseudo-streamer source region.

One major difficulty in constructing more realistic three-dimensional (3D) global MHD models of sympathetic events is that additional assumptions are needed to establish how unstable the triggered eruption might be. However, these assumptions are very hard to constrain from the available observations. Therefore, instead of building a single realistic case, we build a realistic background coronal environment for 2011 February 15 and investigate quantitatively how the eruption of flux ropes having various strengths and orientations might impact the magnetic structures in the near and remote neighborhood of the eruption. With this study, we can achieve a better understanding about the role of large-scale magnetic coupling during solar sympathetic events and can thereby explore several candidate mechanisms for solar sympathy.

The paper is organized as follows: In Section 2, we describe the global MHD model Alfv\'{e}n Wave Solar Model (AWSoM) used to construct the background solar wind and Gibson-Low (GL) flux rope model for the CME initiation. In Section 3, we describe the methods that are used for analyzing the simulation data in this study. The simulation results are shown in Section 4, followed by the discussion and summary in Section 5.

\section{Model}
In this study, we utilize the Space Weather Modeling Framework (SWMF) developed at the University of Michigan, which provides a high-performance computational capability to simulate the space weather environment from the upper solar chromosphere to the Earth's upper atmosphere and/or the outer heliosphere \citep{toth05, toth12}. The framework contains several components that represent different physical domains of the space environment and each physics domain has several models available. We will mainly focus on the Solar Corona (SC), Inner Heliosphere (IH), and Eruptive Event generator (EE) components. For SC/IH, the Block-Adaptive-Tree-Solarwind-Roe-Upwind-Scheme (BATS-R-US) code plays a central role in solving the MHD equations that describe the plasmas in the heliosphere \citep{powell99}.

\subsection{Alfv\'{e}n Wave Solar Model}
The SC model used in this study is the newly developed AWSoM \citep{bart14}, which is a data-driven model with a domain extending from the upper chromosphere to the corona and heliosphere. A steady-state solar wind solution is obtained with the local time stepping and second-order shock-capturing scheme \citep{toth12}. In order to construct a realistic coronal environment, the inner boundary condition of the magnetic field is specified by a global magnetic map sampled from an evolving photospheric flux transport model \citep{schrijver03}. The inner boundary conditions of electron and proton temperature $T_{e}$ and $T_{i}$ and number density $n$ are assumed at  $T_{e}=T_{i}=$ 50,000 K and $n =$2$\times$10$^{17}$ m$^{-3}$, respectively. The overestimated density at the inner boundary allows chromospheric evaporation to self-consistently populate the upper chromosphere with an appropriately high density, as found on the Sun. The inner boundary density and temperature do not otherwise have a significant influence on the global solution \citep{lio09}. The initial conditions for the solar wind plasma are specified by the Parker solution \citep{parker58}, while the initial magnetic field is based on the Potential Field Source Surface (PFSS) model with the Finite Difference Iterative Potential Solver (FDIPS, \citealt{toth11}).

Alfv\'{e}n waves are driven at the inner boundary with the Poynting flux scaling with the surface magnetic field. The solar wind is heated by Alfv\'{e}n wave dissipation and accelerated by thermal and Alfv\'{e}n wave pressure. Electron heat conduction (both collisional and collisionless) and radiative cooling are also included in the model. These energy transport terms are important for self-consistently creating the solar transition region. In order to produce physically correct solar wind and CME structures, such as shocks, the electron and proton temperatures are treated separately \citep{chip12, jin13}. Thus, while the electrons and protons are assumed to have the same bulk velocity, heat conduction is applied only to the electrons, owing to their much higher thermal velocity. By using physically consistent treatment of wave reflection, dissipation, and heat partitioning between the electrons and protons, the AWSoM showed the capability to reproduce the solar corona environment with only 3 free parameters that determine Poynting flux ($S_A / B$), wave dissipation length ($L_\perp\sqrt{B}$), and stochastic heating parameter ($h_S$) \citep{bart14}.

The SC model uses a 3D spherical grid from 1 R$_{\odot}$ to 24 R$_{\odot}$. The grid blocks consist of 6$\times$4$\times$4 mesh cells. The smallest radial cell size is $\sim$10$^{-3}$ R$_{\odot}$ at the Sun, allowing the steep density and temperature gradients in the upper chromosphere to be resolved. The largest radial cell size at the outer boundary of SC is $\sim$1 R$_{\odot}$. Below $r=1.7$ R$_{\odot}$, the angular resolution is $\sim$1.4$^{\circ}$. Above this radius, the grid coarsens by one level to $\sim$2.8$^{\circ}$. The IH model uses a Cartesian grid to reach 250 R$_{\odot}$ with grid blocks consisting of 4$\times$4$\times$4 mesh cells. The smallest cell size in IH is $\sim$10$^{-1}$ R$_{\odot}$ and the largest cell size is $\sim$8 R$_{\odot}$. For both the SC and IH, adaptive mesh refinement (AMR) is performed to resolve the heliospheric current sheet (HCS). The number of total cells is $\sim$3$\times$10$^{6}$ in SC, and $\sim$1$\times$10$^{6}$ in IH. In steady-state, both the SC and IH domains are in heliographic coordinates (rotating at the Carrington rotation rate).

\subsection{Gibson-Low Flux Rope Model}
In this study, we initiate CMEs using the analytical Gibson-Low (GL; \citealt{gibson98}) flux rope model implemented in the Eruptive Event Generator Gibson Low (EEGGL) module. This flux rope model has been successfully used in numerous studies modeling CMEs (e.g., \citealt{chip04a, chip04b, lugaz05a, lugaz05b, schmidt10, chip14}). Analytical profiles of the GL flux rope are obtained by finding a solution to the magnetohydrostatic equation $(\nabla\times{\bf B})\times{\bf B}-\nabla p-\rho {\bf g}=0$ and the solenoidal condition $\nabla\cdot{\bf B}=0$. This solution is derived by applying a mathematical stretching transformation $r\rightarrow r-a$ to an axisymmetric, spherical ball of twisted magnetic flux $\bf b$ with diameter $r_0$ centered relative to the heliospheric coordinate system at $r=r_1$. The field of $\bf b$ can be expressed by a scalar function $A$ and a free parameter $a_1$ to determine the magnetic field strength. The full derivation of $\bf b$ can be found in the Appendix of \citet{lites95}. Following the transformation, the GL flux rope field takes the form:
\begin{equation}
{\bf B_{\rm GL}} (r,\theta,\phi) = \left(\frac{\Lambda}{r}\right)^2 b_r (\Lambda, \theta, \phi) {\bf e_r} + 
\left(\frac{\Lambda}{r}\frac{d\Lambda}{dr}\right) b_\theta(\Lambda, \theta, \phi) {\bf e_\theta} +
\left(\frac{\Lambda}{r}\frac{d\Lambda}{dr}\right) b_\phi(\Lambda, \theta, \phi) {\bf e_\phi}
\end{equation}
where $\Lambda = r + a$. Equilibrium for the transformed state requires the plasma pressure as:
\begin{equation}
p_{\rm GL} = \left(\frac{\Lambda}{r}\right)^2\left(1-\left(\frac{\Lambda}{r}\right)^2\right)\left(\frac{b_{r}^2}{2\mu_0}\right)+\left(\frac{\Lambda}{r}\right)^2 a_1 A
\end{equation}
and the density be of the form:
\begin{equation}
\begin{split}
\rho_{\rm GL} &= \frac{1}{F(r)}\left[-\left(\frac{\Lambda}{r}\right)^2\left(1-\left(\frac{\Lambda}{r}\right)^2\right)\frac{d}{d\Lambda}\left(a_1 A + \frac{b^2}{2\mu_0}\right)+ 2 a_1 A \frac{\Lambda a}{r^3} \right.\\ 
&+ \left. \frac{\Lambda a}{\mu_0 r^3}\left(1-2\left(\frac{\Lambda}{r}\right)^2\right) b_{r}^2
+\left(\frac{\Lambda}{r}\right)^2\left(\frac{a^2}{r^2}+\frac{2a}{r}\right) \left(\frac{b_{\theta}^2+b_{\phi}^2}{\mu_0\Lambda}\right)\right]
\end{split}
\end{equation}
where $F(r)=GM/r^2$, $G$ is the gravitational constant and $M$ is the solar mass. The transformed flux rope appears as a tear-drop shape of twisted magnetic flux. At the same time, Lorentz forces are introduced, which leads to a density-depleted cavity in the upper portion and a dense core at the lower portion of the flux rope. This flux rope structure helps to reproduce the 3-part density structure of the CME in the observation \citep{llling85}. However, the dense core in the GL flux rope is not highly structured as is observed. The GL flux rope and contained plasma are then superposed onto the steady-state solar corona solution: i.e. $\rho=\rho_{0}+\rho_{\rm GL}$, ${\bf B = B_{0}+B_{\rm GL}}$, $p=p_{0}+p_{\rm GL}$. The combined background-flux rope system is in a state of force imbalance (due to the insufficient background plasma pressure to offset the magnetic pressure of the flux rope), and thus erupts immediately when the numerical model is advanced forward in time. There are several advantages of using this unstable flux rope for this study. First, the eruption does not require time for energy to build up that is computationally expensive. Second, the GL flux rope enables us to estimate the total magnetic energy added to the system, therefore facilitates a parameter study.

The initial out-of-equilibrium flux rope eruption does have some artifacts that should be noted. Due to the force imbalance, the initial acceleration process in the source region may not be captured correctly in the simulation. In this study, we focused on the dynamic evolution of surrounding solar structures rather than the eruption itself. The initial arbitrary acceleration in the source region therefore has limited influence on these structures due to the long distance and the cumulative effect of magnetic reconnection between the erupting flux rope and the ambient fields. However, we noticed that the strong acceleration in the beginning may lead to stronger fast-mode waves/shock-waves than those in the realistic events. Therefore, the impact caused by waves (see \S 4.2.2) may be overestimated in this study. To overcome this drawback, a more self-consistent modeling of pre-eruptive configurations is needed (e.g., \citealt{titov14}).

The GL flux rope is mainly controlled by four parameters: the stretching parameter $a$ determines the shape of the flux rope, the distance of torus center from the center of the Sun $r_{1}$ determines the initial position of the flux rope before it is stretched, the radius of the flux rope torus $r_{0}$ determines the size of the flux rope, and the flux rope field strength parameter $a_1$ determines the magnetic strength of the flux rope. The location (longitude and latitude) and orientation of the flux rope are also specified in the simulation. For this study, we fix three parameters ($a=0.3$, $r_{0}=0.3$, $r_{1}=1.4$) and location of the flux rope. We conducted a total of 11 runs with different flux rope $a_1$ and orientations. The GL flux rope parameters used in this study are shown in \S 3.

\section{Method}
\subsection{Reconstruction of the Solar Corona}
By using the global map, as taken from the time-evolving surface flux transport model of \citet{schrijver03} on 2011 February 15 00:04:00 UT, and running AWSoM in local time-stepping mode that allows fast convergence, a steady-state MHD solution of the solar corona at that time can be obtained. For a complete list of model parameters of the AWSoM, refer to the Table 1 of \citet{bart14}. Here, we use all parameters except $(S_A/B)_\odot$, which controls the Poynting flux passing through the surface of the Sun. A larger value of this parameter increases the level of coronal heating and therefore opens up more field into the heliosphere, leading to larger coronal holes in the synthesized extreme ultraviolet (EUV) images. We chose $(S_A/B)_\odot= 1.2\times10^6$ W m $^{-2}$ T$^{-1}$ to get a better match between the synthesized and observed EUV images near the Sun. In Figure 1, the model density, temperature, and the instrumental response functions constructed from the CHIANTI 7.1 atomic database \citep{dere97, landi13} are used to synthesize EUV images, which are then compared with the EUV observations from SDO/Atmospheric Imaging Assembly (AIA; \citealt{lemen12}) and STEREO/Extreme UltraViolet Imager (EUVI; \citealt{howard08}). Three EUV spectral bands (AIA 211~\AA\, EUVIA 171~\AA\, and EUVIB 195~\AA) are selected that cover the temperature range from 1 MK to 2 MK. It is evident in the figure that emission from all the major active regions and coronal holes is reproduced in the synthesized AIA 211~\AA\ image. However, in the synthesized STA/STB images, several active regions are missing due to the outdated far-side magnetic map (STA was $\sim$87$^\circ$ ahead of Earth and STB was $\sim$94$^{\circ}$ behind Earth at that time). Identical log scales are used for both the observed and synthesized images.

In order to further validate the steady-state background solar wind, we also compare the modeled unsigned openflux with that measured at 1 AU by the \textit{in situ} OMNI database\footnote{The OMNI database (obtained from the National Space Science Data Center (NSSDC)) provides selected data from the Advanced Composition Explorer (ACE), Wind, Geotail, and IMP8 spacecraft (IMP8 ceased operation on October 7, 2006).}. Under the assumption of constant interplanetary magnetic field (IMF) $|B_{r}|$ in latitude \citep{suess96a, suess96b}, the total unsigned openflux is calculated as $F_{1AU}=4\pi r^2 |B_{r}| $, where  $|B_{r}|$ is averaged over one Carrington rotation. The unsigned openflux in the database is calculated to be $5.73\times10^{22}$ Mx, while the unsigned openflux in the model is $5.33\times10^{22}$ Mx.

In Figure 2, we compare the initial potential field (blue) with the relaxed, steady-state MHD solution (red) from AWSoM. Carrington coordinates are used for all the 3D simulation data in this study. We select the same foot points for both the potential and MHD field lines in the figure. In the left panel, the closed field near the Sun is shown with the isosurface showing the current density $|{\bf J}|=2\times10^{-7}$ A m$^{-2}$. In general, the PFSS and MHD solutions are very similar for small loops, but the larger loops in the MHD solution are more radially stretched. This result is consistent with a study by \citet{riley06}. In the right panel, the helmet streamer belt in the two solutions may be compared. The helmet streamer field lines are selected according to the HCS locations at 2.5 R$_{\odot}$ for PFSS solution and 3.0 R$_{\odot}$ for MHD solution. In addition to being stretched to a higher altitude, the MHD helmet streamer belt has also reached higher latitudes at certain places than the corresponding potential field solution.

\subsection{Quantifying the Eruption Impact}
A GL flux rope located within active region (AR) 11158 is inserted into the model. The location is marked as ``CME" in Figure 3. The insertion of the flux rope adds a mass of 2.4$\times$10$^{16}$ g and a magnetic energy 4.1$\times$10$^{32}$ ergs to the solar corona\footnote{This calculation is based on the Run 1 flux rope parameters in Table 1.}, both of which are typical parameters associated with X-class flare/CME systems (e.g., \citealt{emslie05}). The simulation is evolved forward in time for 1 hour after flux rope insertion, during which time CME passes through the model corona. The other active regions and filament channels on disk at the time of the eruption are shown in Figure 3. In the following, we briefly describe these solar structures, quantify the eruption impact, and summarize the simulation runs. 

\subsubsection{Selection of the Solar Structures}
The objective is to investigate the impact of a CME on all the active regions (with or without an AR number), filament channels, and some quiet Sun regions on disk at the time of the liftoff of the CME. In total, there are 8 active regions, 5 filament channels, and 2 quiet Sun regions, all of which are labeled in the left panel of Figure 3. The right panel of Figure 3 shows the H$_{\alpha}$ observation, in which the filament channels are evident. Note that the filament channels in this simulation do not contain flux ropes. Instead, they are modeled by the MHD solution with near-potential fields. Therefore, the filament channels in this study are essentially referred to diffuse region polarity inversion lines (PILs). 

\subsubsection{Decay Index}
To characterize the impact of the eruption on the various structures, we evaluate the decay index $I_{\rm Decay}$ before and after the eruption:
\begin{equation}
I_{\rm Decay}=-\frac{d\log B}{d\log h}
\end{equation}
where $B$ is the magnetic field strength and $h$ is the height above the solar surface. The decay index represents how fast the overlying field decays with height, which is a key factor for determining when the instabilities happen. Faster decay corresponds to a lesser confining force, which in turn makes eruptions more likely. The critical decay index $I_{crit}$, above which the flux rope becomes unstable, depends on the flux rope configuration: 1.0 for a straight line current \citep{vantend78} and 1.5 for a toroidal current \citep{bateman78}. Several observational studies find that active regions with a larger decay index are more likely to generate CMEs (e.g., \citealt{liu08}). In a statistical study, \citet{filippov08} find that source regions for erupted filaments have a larger decay index than those for stable filaments. Theoretical and numerical studies suggest that a critical decay index between 1.0 and 2.0 is a good approximation for solar corona conditions under various assumptions (e.g., \citealt{kliem06, aulanier10, demoulin10}). In this study, we extract the magnetic field profile above the PIL of different solar structures. We then calculate the decay index for the selected solar structures between 5 Mm and 100 Mm and track its evolution after the eruption, from which we characterize to what extent the solar structures are affected by the CME.

\subsubsection{Impact Factor}
Because the plasma parameters (e.g., field, velocity, pressure, density) may undergo dramatic changes during the eruption, the decay index alone may not fully characterize the CME's impact magnitude on different structures. In order to quantify the impact more comprehensively, we introduce the following two impact factors. The first is the sum of relative perturbations in relevant physical quantities:
\begin{equation}
F_{{\rm imp1},i} = \frac{1}{7}\cdot\left( \Delta\rho_{i} + \Delta{B}_{i} + \Delta P_{i} + \Delta F_{{\rm Lorentz},i} + \Delta U_{i} + \Delta I_{{\rm Decay},i} + \Delta I_{{\rm Final},i} \right)
\end{equation}
where $\rho$, $B$, $P$, $F_{\rm Lorentz}$, and $U$ represent density, total magnetic field, total pressure, Lorentz force, and total velocity, respectively. $\Delta\rho_{i}=\displaystyle \max_{t=0}^{1hr} \left |\frac{\Delta\rho_{t}}{\rho_{0}} \right |_{i}$, $\Delta B_{i}$, $\Delta P_{i}$, $\Delta F_{{\rm Lorentz}, i}$, $\Delta U_{i}$, and $\Delta I_{{\rm Decay}, i}$ have similar definitions, while $F_{{\rm Lorentz}}=| {\bf J} \times {\bf B} |/ B^2$. $\Delta I_{{\rm Final}}=\left | \frac{I_{{\rm Decay},1hr}-I_{{\rm Final}}}{I_{{\rm Final}}} \right |$, where $I_{{\rm Final}}$ refers to the decay index derived from the relaxed steady-state solution calculated from the modified inner boundary map, which is comprised of the global magnetic map plus the contribution of the GL flux rope. This term describes how close the decay index (at t = 1 hour) is to the final relaxed MHD solution. The impact factor $F_{\rm imp1}$ thus takes into account changes in 7 different impact quantities over the course of the first hour of evolution. All the valuables are averaged between 5 and 100 Mm, and are weighted equally.

The second impact factor has a similar definition but is normalized to the maximum relative impact seen within the 15 selected regions. Also a new quantity (penetration height) is added:
\begin{equation}
\begin{split}
F_{{\rm imp2}, i} =\frac{1}{8}\cdot &\left(\frac{\Delta\rho_{i}}{\displaystyle \max_{i=1}^{15} \Delta\rho_{i}} + \frac{\Delta B_{i}}{\displaystyle \max_{i=1}^{15} \Delta B_{i}} +\frac{\Delta P_{i}}{\displaystyle \max_{i=1}^{15} \Delta P_{i}} + \frac{\Delta F_{{\rm Lorentz},i}}{\displaystyle \max_{i=1}^{15} \Delta F_{{\rm Lorentz},i}} + \frac{\Delta U_{i}}{\displaystyle \max_{i=1}^{15} \Delta U_{i}} \right.\\ 
&+ \left. \frac{\Delta I_{{\rm Decay},i}}{\displaystyle \max_{i=1}^{15} \Delta I_{{\rm Decay},i}} + \frac{\Delta I_{{\rm Final},i}}{\displaystyle \max_{i=1}^{15} \Delta I_{{\rm Final},i}} + \frac{1/H_{i}}{\displaystyle \max_{i=1}^{15} 1/H_{i}}  \right)
\end{split}
\end{equation}
The penetration height $H_{i}$ is defined as the lowest altitude where the magnetic field magnitude changes by at least 10\% relative to the pre-eruption field. This impact factor can vary from 0 (no impact) to 1 (largest impact among structures). As we can see from the equations, the first impact factor is useful when comparing the impact amongst the set of simulation runs with different initial flux rope parameters. The second impact factor is useful when comparing the impact of the CME on the various structures in the same simulation run.

\subsection{Summary of the Simulation Runs}
In Table 1, we summarize the various simulation runs, in which the calculations were initialized with different GL flux rope parameters. The two major parameters that we experimented with are the initial orientation angle and the magnetic field strength of the flux rope.  As shown in Table 1, for Runs 1--7, the flux rope orientation angle various between 90$^{\circ}$ and 270$^{\circ}$, allowing the study of the dependence of the impact on orientation. For Runs 1, 8, 9, and 11, only the magnetic field strength of the flux rope is varied, which enables the energy dependence of the impact to be assessed. The convention used here is that the an orientation angle of 0$^{\circ}$ means that the foot points of the flux rope are along the East-West direction with the positive polarity at East, while an orientation angle of 180$^{\circ}$ has the positive polarity to the West. The orientation angle increases in a clockwise fashion.

\section{Results}
\subsection{Dynamic Evolution after the Eruption}
The corona evolves dramatically after the eruption. In Figure 4, we show the plasma-$\beta =2\mu_{0}\cdot(p_{e}+p_{i})/B^2$ at 2.5 R$_{\odot}$ (a-c) and radial velocity at 42 Mm (d-f) for t = 0, 10, and 60 min, respectively. In general, the high plasma-$\beta$ regions at 2.5 R$_{\odot}$ approximate the HCS location. From the evolution of the plasma-$\beta$, we can see clearly that the eruption changes the large-scale magnetic configuration, and thus the HCS, significantly. After an hour of evolution, the HCS has still not relaxed to the original state at t = 0. The major changes occur around $\pm$50$^\circ$ in longitude around the CME source region (AR 11158). In that region, the HCS locations are pushed to higher latitudes both in the north and south hemispheres. 

In Figure 4(d)-(f), we show the radial velocity evolution at 42 Mm, which approximately resembles the height of the Dopplergram observed by spectral lines with log T = 5.5 -- 6.0 \citep{guo09}. At t = 10 min, we can see clearly the strong upward motion inside the expanding CME ``bubble", in front of which the downward motion is evident with a maximum value of $\sim$100 km s$^{-1}$ (Figure 4(e)). The downward motion is caused by the downward push of the CME during its expansion into the corona. This same phenomenon was observed to occur in AR 11158 by \citet{harra11} and \citet{veronig11} using data from the Hinode/EUV imaging spectrometer (EIS; \citealt{culhane07}) during an M1.6 flare/CME event on 2011 February 16. The upward motion is induced by the radial propagation of the CME and is more often observed in the Dopplergrams after CMEs (e.g., \citealt{harra07, imada07, jin09, tian12}). At t = 1 hour, the plasma speed has mostly decreased back to the pre-event magnitude, except for a small area around the CME source region where some upward motion is still evident ($\sim$70 km s$^{-1}$).

In order to illustrate the global magnetic configuration change during the eruption, we further show in Figure 5 the 3D field configuration at t = 0 (left column) and 15 min (right column) from three different points of view (an animation is available online). The lines represent flux rope field lines (red), large-scale helmet streamers (white), and field lines from surrounding active regions/open field (green). At t = 15 min, the flux rope structure is a mix of three different ensembles (colors) of  field lines due to magnetic reconnection. Also, the large-scale helmet streamers around the CME flux rope are significantly disturbed, either by field line compression or reconnection.

\subsection{Different Types of CME Impact}
By analyzing the simulation data using the method described in \S 3, we can identify the factors that determine the CME impact. Ideally, we expect that the CME impact is determined by the distance from the source region as well as by the magnetic field strength of the structures. In Figure 6, we show two scatter plots between the impact factors (defined in \S 3) as a function of the distance (Figure 6(a)) and magnetic field strength (Figure 6(b)) based on the data from Run 1. Larger symbol sizes in the figure reflect stronger magnetic field strengths (Figure 6(a)), whereas in Figure 6(b), larger symbols indicate closer distances to the source region. The magnetic field strength is the average value between 5 Mm and 100 Mm in height. For display purposes, the impact factor $F_{{\rm imp1}}$ is normalized by the largest value. The general trend is that stronger CME impacts occur when the target region has weaker field strength, and especially when the target region is closer to the source region. All regions with impact factor larger than 0.2 are located within 400 Mm from the source region, while these regions can have a broad range of magnetic field strengths. We now discuss additional contributing factors such as magnetic field line connectivity and topology.

The CME impact morphologies can be placed into three categories:
\subsubsection{Type I: direct connection}
This type of impact applies to the solar structures that the flux rope expansion can reach directly. Due to the direct interaction between the flux rope magnetic field and the field of the impacted solar structures, the severity of Type I impacts depends strongly on the initial orientation of the flux rope in the simulation. For the simulated CME erupting from AR 11158 discussed here, we find that regions that experience Type I impacts include QR1, AR2, FC1, FC2, and AR8 (marked with red in Figure 3). In Figure 7, we show as an example the Type I impact for AR2 (an animation is available online), in which we compare the magnetic field configuration as well as the decay index with height for two different flux rope orientations (Run 1 and 2). 

At around 15 min after the eruption, the CME expansion reaches AR2 and starts to interact with the AR2 magnetic field. Figures 7(a) and 7(b) show the global magnetic field configuration at t = 15 min with selected field lines to represent flux rope (red), large-scale helmet streamers (white), and field lines from surrounding active regions and open field (green). A zoom-in view of AR2 can be found in Figure 7(c) and 7(d) with the background showing the normalized Lorentz force and the coloring of the field lines showing the height information. In order to distinguish the field line connectivity, the field lines that reconnect with the erupting flux rope are shown in white color. For Run 1, we can see that the flux rope expansion induces more magnetic reconnections than the expansion in Run 2, as indicated by the larger fraction of field lines connecting AR2 and the CME source region in Run 1. As a consequence, the reconnection in Run 1 removes more overlying field from AR2 and leads to a higher decay index with height (Figure 7(e) and 7(f)). Note that CME-induced reconnection has been observed and simulated by \citet{van14}. In the decay index figure, we also show the PFSS solution as well as MHD final solution (new steady-state after flux rope insertion) for comparison. For AR2, the PFSS and MHD final solutions overlap.

In Figure 8, another example of a Type I impact is shown for filament channels FC1 and FC2, which are located southwest of the CME source region. Due to the weaker magnetic field strength above the filament channel ($\sim$20 G at 5 Mm) compared with AR2 ($\sim$35 G at 5 Mm), the CME has a larger and more substantial impact on these structures. Again, we compare the results for two different flux rope orientations (Run 1 and 2). In contrast to AR2, the impact on FC1 and FC2 is larger in Run 2 than in Run 1. As shown in the field line configurations shown in Figure 8(a) and 8(b), we find that there are more open field lines created in Run 2 than in Run 1 for both FC1 and FC2 at t = 30 min, which in turn leads to a greater change in the overlying field strength and therefore a larger decay index. Note that the decay index starts to decrease above $\sim$50 Mm for FC1 in the Run 2 case, an effect likely caused by the dramatically lower magnetic field strength below 50 Mm. Also, we notice that the decay index change for FC2 above 80 Mm is bigger for Run 1 than Run 2. For FC1 and FC2, we can see that the potential and MHD solutions are slightly different, with larger differences at higher altitude.
 
In order to further determine the magnitude of the impact as the flux-rope orientation changes, and to characterize when the impact is largest for different solar structures, we conduct a total of 7 simulation runs (Run 1--7) with different flux rope orientations, while keeping the other flux rope parameters fixed, and calculate the impact factor $F_{{\rm imp1}}$ for the solar structures in each run. Figure 9 shows the orientation dependence of CME impact on two structures with type I impacts (AR2 and AR8), and indicates that the magnitude of impact changes significantly in AR2 and AR8 with different flux rope orientations. For AR2, the difference between the largest and smallest impact factor is $\sim$30\%, whereas for AR8 the difference is $\sim$40\%. In the left panel of Figure 9, we mark the largest impact configuration for both AR2 (red) and AR8 (blue). The arrows in AR2 and AR8 show the orientation of the active region (pointing from positive to negative polarity), while the arrows in the CME source region show the orientations of the flux rope with the largest impact. The results shown here suggest that the largest impact occurs when the CME flux rope orientation favors reconnection with the impacted solar structures during the expansion phase.

In the numerical simulations, the interaction between the flux rope and the surrounding magnetic structure is traced through 3D field evolution, however such information cannot be easily obtained from observations. To make the simulation results more readily comparable to observations, we now use the simulation data to synthesize time series of EUV images. These synthesized images clearly reveal EUV waves, namely bright fronts that propagate on a global scale, as seen in EUV images from EIT and AIA. Note that the EUV wave in the 2011 February 15 event was studied by \citet{schrijver11b}, \citet{olmedo12}, and \citet{nitta13a}.  Although there is some debate about the origin of EUV waves, there seems to be a consensus that a single event could consist of both the ``wave" (fast MHD wave) and ``non-wave" (CME structure) components (see recent reviews by \citealt{chen05}, \citealt{pat12}, \citealt{liu14}, and \citealt{warmuth15}).  Which part of the front represents which processes remains unclear, however.  This is where numerical experiments may help reduce the ambiguities of the observations.  

We try different flux rope orientations and evaluate the properties of the ensuing EUV waves. For this purpose, we select two simulation runs with moderate flux rope energy and initial orientation of 128$^\circ$ and 270$^\circ$ (Run 9 and 10). A moderate flux rope energy is chosen to approximately match the intensity enhancement of EUV waves in the observation, and the two orientation angles were chosen to correspond to the maximum and minimum impact factors for AR2 shown in Figure 9. Tri-ratio EUV images were then synthesized by dividing two subsequent simulated images in selected wavelengths at times with an appropriate time interval between them (t = 6 min / t = 4 min) for both runs, as shown in the left and middle panels of Figure 10. The images are composites of AIA 211~\AA\ (red), AIA 193~\AA\ (green), and AIA 171~\AA\ (blue). The ratio in each channel is scaled to 1$\pm$0.3. Because these three filters have the highest signal-to-noise ratio on AIA, the tri-ratio method serves as a useful tool to investigate EUV waves (e.g., \citealt{downs12}) and eruptions (e.g., \citealt{nitta13b}). Comparing the synthesized images in the two runs indicates that the outer front of the EUV waves (marked by blue arrow) is quite similar for the two orientations, while certain areas of the inner front (marked by red arrow) show a marked difference. 

These differences can be interpreted with the assistance of 3D field configuration shown in the right panel of Figure 10. The outer front has a fast-mode wave nature driven by the eruption, which has been simulated in previous studies (e.g., \citealt{wu01, cohen09, downs11, downs12}). The intensity increase in the EUV bands is caused by adiabatic compression, which can be seen as a density increase region in the 3D configuration image (red surface). The inner front represents the expanding volume of the flux rope, where is the main site for reconnection between the flux rope and the surrounding fields. This wave/non-wave nature of EUV waves has also been found in the previous numerical studies \citep{cohen09, downs12}. Behind that front is a low density region (blue surface) with most of the flux rope field lines contained inside. The intensity difference of the inner front near AR2 found in the two runs is due to magnetic reconnection between the flux rope and AR2 field. When the flux rope orientation is more favorable to reconnection with AR2, as in Run9, the intensity increases. This effect may be used to identify the reconnection site in observations, and may also help to constrain the orientation of the expanding flux rope. Our result suggests that these numerical models may help us diagnose the origin of the slow and late EUV fronts that brighten due to magnetic reconnection \citep{guo15}.

\subsubsection{Type II: indirect connection}
This type of impact includes solar structures that the flux rope cannot reach directly. In contrast to the type I impact, we find that Type II impacts do not significantly depend as much on the orientation of the flux rope. In this study, the structures with Type II impacts include AR1, AR3, AR4, AR5, FC3, FC4, FC5, and QR2 (marked with green in Figure 3). As an example, we show the orientation dependence of the CME impact on AR1 in Figure 9. In contrast to  Type I impacts, the calculated impact factor does not depend on the flux rope orientation, and thus does not involve pronounced magnetic reconnection. Instead, the impact is usually caused by field line compression during the eruption (i.e., fast-mode waves). Because no reconnection is involved, the decay index decreases slightly during the compression process and relaxes back to the pre-event state after the wave passage.

Another mechanism that can cause Type II impacts is through the recovery phase\footnote{The recovery phase refers to the evolution after the main phase of the CME impact. The main phase is defined as when the average normalized Lorentz force in the active region reaches maximum in the simulation.} evolution. An example is shown in Figure 11, in which we show the magnetic field evolution of AR5 during the hour after the eruption. We can see that after the main phase of CME impact (t = 8 min), the field lines start to relax back to the original state. However, the field of AR5 starts to expand again after t = 26 min and keeps changing at t = 60 min. This evolution is induced by the large-scale structure changing at higher altitudes during the CME propagation. The field lines above AR5 are stretched during this process, which leads to a larger pressure gradient that causes the lower field to expand again in the recovery phase. Our simulation suggests that although the main phase of CME impact only lasts for several minutes, the post-eruption reconfiguration can last for hours after the eruption. The fact that the expansion in the recovery phase causes an increase in the decay index shows that this post-eruption reconfiguration process could play an important role for some of the solar sympathetic events.

\subsubsection{Type III: hybrid wave-reconnection coupling}
Type III impacts are hybrids between Type I and Type II. Their evolution does depend on the orientation of the flux rope, but the influence is less pronounced than Type I impacts. For example, the effects on AR6 and AR7 (marked with blue in Figure 3), which are not directly connected to the source region, occur as a result of their proximity to AR2, which we classified earlier as having Type I impact. The interaction between Type III structures and the CME often occur via nearby structures possessing Type I impacts. Because the magnetic configuration can be quite different for Type I impacts under different flux rope orientations, this difference may change the magnetic configuration of regions showing Type III impacts as a result. But due to the indirect nature, the influence of orientation is less evident than for the Type I impacts. In Figure 9, we show the orientation dependence of the CME impact on AR6 as an example.

\subsection{Energy Dependence}
Another factor that may influence CME impacts is the energy of the eruption. In order to investigate the energy dependence of the CME impact on different structures, we change $a_1$ parameter in Equation (2) and (3) while keeping the other parameters fixed. Run 1, 8, 9, and 11 possess four different energy inputs, covering a range spanning a factor of about 40: 1.13$\times$10$^{31}$, 4.12$\times$10$^{31}$, 1.23$\times$10$^{32}$, and 4.06$\times$10$^{32}$ ergs. For each run, we calculate the CME impact on the selected set of structures, as shown in Figure 12. As expected, the magnitude of impact generally increases for larger energy input. An interesting finding is that for most of the structures with Type II impacts, the magnitude of the impact and energy input have a quasi-linear relationship, while for Type I and III structures, that relationship is non-linear. According to the characteristics of different impact types mentioned in \S 4.2, it suggests that environments that facilitate magnetic reconnection between two regions cause the impact factor to depend nonlinearly on the energy input. If no reconnection is involved in the process, the impact magnitude increases linearly with the CME energy. We also find that all the remote regions feature the same impact types across all energies in this study.

We also try to relate the CME speeds with the corresponding impacts in the simulations. The resulting CME speed for each energy input is shown in Figure 12 in the unit of km s$^{-1}$. The CME speed is defined as the average speed of the outmost density enhancement front between 30 and 60 minutes in the simulation. The CME speeds in our simulations range from 411 km s$^{-1}$ to 2607 km s$^{-1}$. We also found that the square of the CME speed has a linear relationship with the magnetic energy of the GL flux rope.

\subsection{Influence of Field Topology}
In previous sections, the connectivity between the eruptive flux rope and the various target features is determined by tracing and visualizing field lines associated with the unstable flux rope (e.g., in Figure 5 and 7). However, it is unclear how the boundary of the flux rope expansion is determined. Analyzing the large-scale field topology provides additional insight (e.g., \citealt{longcope05}). In this study, we assume that the PFSS solution could be a good estimation for the large-scale configuration of the corona field, which is presumed to be mostly relaxed and thus close to the potential state. In Figure 13, we show the topological structures calculated from the PFSS model, in order to better understand the magnetic environment into which the CME flux rope propagates. Figure 13(a) shows all null points, spine lines, and separatrix surfaces from the PFSS field, while Figure 13(b) shows only those topological domains that directly connect to the CME source region. The methods used to calculate these topological structures are described by \citet{haynes07, haynes10}, after adapting for spherical geometrics. The very large yellow semi-transparent surface demarcates the boundary between the open and closed flux in the PFSS model. The separatrix surfaces evident underneath are topological domains associated with null points in the coronal volume. The blue line shows the null line on the source surface, and approximates the location of the base of the HCS. A detailed description of these topological features can be found in \citet{platten14}. 

We find that all the Type I impacts identified in the simulations occur within topological domains that have direct connections with the CME source region. Considering the similarity between the PFSS and MHD field solution in this study (shown in Figure 2), this topology analysis suggests that the erupting flux rope does not easily break out of the domain of self-contained field from which it originates. PFSS models seem useful for identifying the different types of impacts on solar structures caused by an erupting flux rope, however, additional regions need to be analyzed in this way in order to more firmly establish this possibility.

\section{Summary \& Conclusion}
We have constructed a global solar coronal model for 2011 February 15, with the aim of analyzing the effects of a flux-rope eruption from AR 11158 on the surrounding features visible on the solar disk. Unstable CMEs having different flux rope parameters (e.g., orientation, strength) were inserted into this model, allowing the dependence of the eruption on these parameters to be investigated. The main conclusions are:

1. The impact of a solar eruption on the surrounding solar structures depends on the distance and the magnetic strength of the impacted structures, as well as on the presence (or absence) of a direct coupling mechanism between the CME flux rope and the surrounding large-scale magnetic field. Within the CME expansion domain where the CME flux rope field directly interacts with the solar structures, expansion induced reconnection effectively weakens the overlying field, leading to an increase of decay index. This mechanism may be responsible for coupled eruptions in certain solar sympathetic events. The magnitude of the impact is found to depend on the orientation of the erupting flux rope, with the largest impact occurring when the CME flux rope is favorably oriented for reconnecting with the surrounding regions. Moreover, the magnitude of the impact appears to increase more weakly than linearly with eruption energy.

2. Outside the CME expansion domain, the influence of the CME is mainly through field compression by fast-mode waves, with a magnitude that is roughly proportional to eruption energy. Because no direct reconnection is involved, the decay index in the low corona always decreases during the wave passage and largely relaxes back to the pre-event state thereafter. Even with the energy input of an X-flare/CME, the waves by themselves have limited impact on a distant active region.

3. For certain structures outside the CME expansion domain, the influence of CME can also occur through the post-eruption reconfiguration of the large-scale field that can persist for hours. During this process, although the major impact of the CME has passed, the magnetic field over the structure continues to evolve, causing the decay index to slowly increase. Therefore, it could be an important factor for certain solar sympathetic events.

Based on the results of this numerical study, we can summarize a list of factors that may determine the CME impact on the different solar structures: the distance from the source region, overlying field of the structure, the relative position of the structure to the source region, the relative orientation of source region and impacted structure, the energy of the eruption, and the topology of the large-scale magnetic field. All of the above-mentioned factors are measurable or can be derived from observations. Therefore, it appears possible to establish an empirical relationship to describe the CME impact in future studies. If such an empirical relationship were found, it could be used to determine regions that will experience the greatest influence even before the actual eruption happens,  which is potentially useful for the purpose of space weather forecasting.

\begin{acknowledgements}
We are very grateful to the referee for constructive comments that helped to improve the paper. M.Jin is supported by UCAR/NASA LWS Jack Eddy Postdoctoral Fellowship. NVN has been supported by NSF grant AGS-1259549. M.Jin is grateful for the hosting of Lockheed Martin Advanced Technology Center. M. Jin thanks W. Manchester, B. van der Holst, W. Liu, T. T{\"o}r{\"o}k, and D. Longcope for invaluable discussions. The simulation results were obtained using the Space Weather Modeling Framework (SWMF), developed at the Center for Space Environment Modeling (CSEM), University of Michigan. We are thankful for the use of the NASA Supercomputer Pleiades at Ames and its helpful staff for making it possible to perform the simulations presented in this paper. We also acknowledge the Texas Advanced Computing Center (TACC) at The University of Texas at Austin for providing HPC resources that have contributed to the research results reported within this paper.

SDO is the first mission to be launched for NASA's Living With a Star (LWS) Program. STEREO (Solar TErrestrial RElations Observatory) is the third mission in NASA's Solar Terrestrial Probes program (STP). The STEREO/SECCHI data are produced by a consortium of NRL (U.S.), LMSAL (U.S.), NASA/GSFC (U.S.), RAL (UK), UBHAM (UK), MPS (Germany), CSL (Belgium), IOTA (France), and IAS (France). The OMNI data access is provided by the NASA Goddard Space Flight Center Space Physics Data Facility (SPDF). H$_\alpha$ data were provided by the Kanzelh{\"o}he Observatory, University of Graz, Austria.
\end{acknowledgements}

\newpage

\clearpage
\begin{deluxetable}{ccc}
\tablecolumns{3}
\tablewidth{0pt}
\tabletypesize{\footnotesize}
\tablecaption{Summary of the Simulation Runs}
\tablehead{
\colhead{}    &  \multicolumn{2}{c}{Flux Rope Parameters} \\
\cline{2-3} \\
\colhead{Run Number} & \colhead{$a_1$\tablenotemark{a}} & \colhead{Orientation}}
\startdata
1  &  50.0 & 128$^{\circ}$  \\
2 &  50.0 & 216$^{\circ}$ \\
3 &  50.0 & 90$^{\circ}$  \\
4 &  50.0 & 156$^{\circ}$\\
5 &  50.0 & 246$^{\circ}$\\
6 &  50.0 & 180$^{\circ}$\\
7 &  50.0 & 270$^{\circ}$\\
8 &  25.0 & 128$^{\circ}$\\
9 &  12.5 & 128$^{\circ}$\\
10 &  12.5 & 270$^{\circ}$\\
11 &  5.0 & 128$^{\circ}$\\
\enddata
\tablenotetext{a}{$a_1$ determines the magnetic strength of the flux rope. The other three parameters of the flux rope are fixed in this study ($a=0.3$, $r_{0}=0.3$, $r_{1}=1.4$).}
\end{deluxetable}

\newpage
\begin{figure}[h]
\begin{center}$
\begin{array}{c}
\includegraphics[scale=0.8]{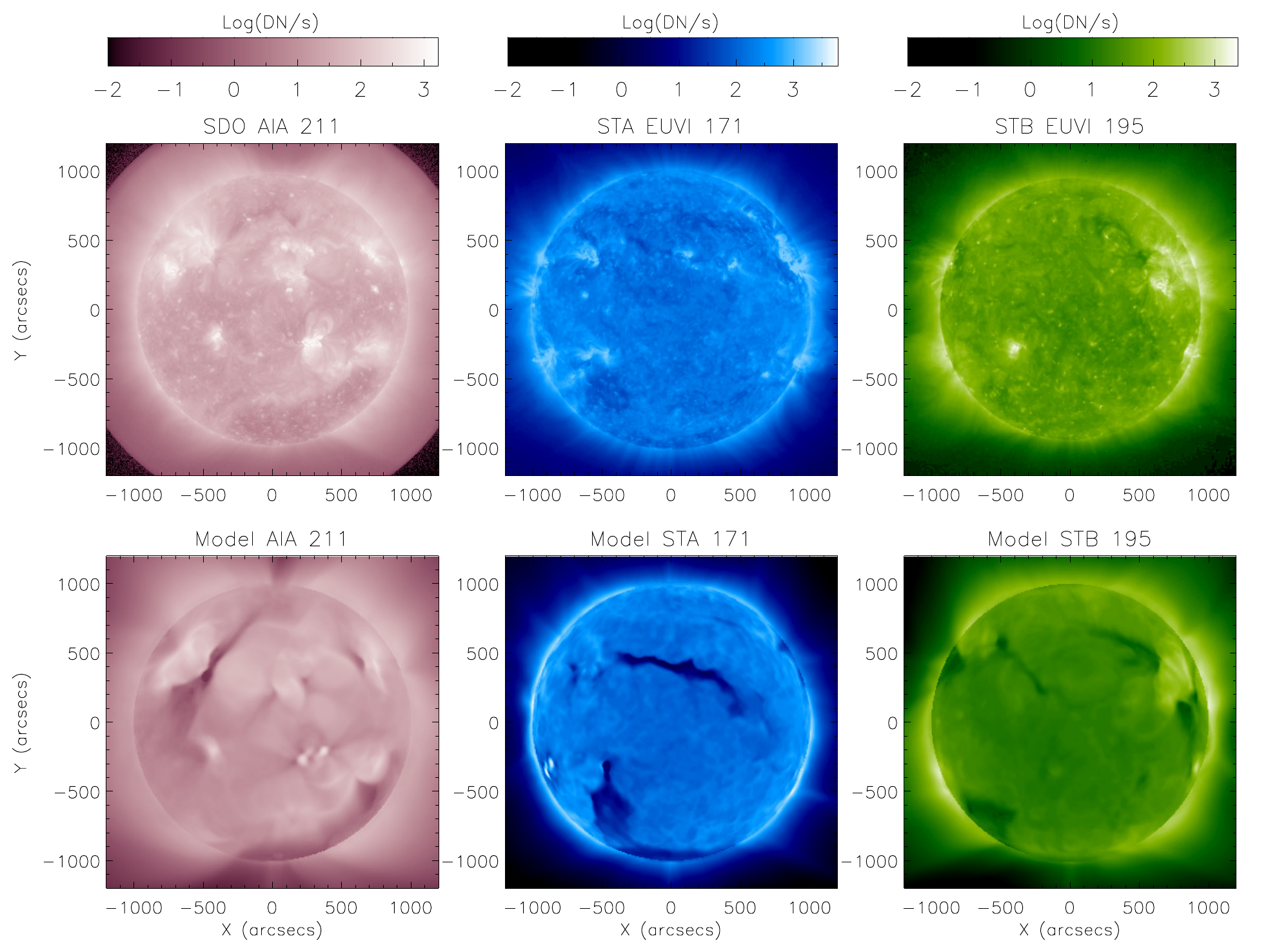}
\end{array}$
\end{center}
\caption{Comparison between observations and synthesized EUV images of the steady-state solar wind model. The observation time is 2011 February 15 $\sim$12:00:00 UT. Top panels: Observational images from SDO AIA 211~\AA, STEREO A EUVI 171~\AA, and STEREO B EUVI 195~\AA. Bottom panels: synthesized EUV images of the model.}
\end{figure}

\newpage
\begin{figure}[h]
\begin{center}$
\begin{array}{cc}
\includegraphics[scale=0.21]{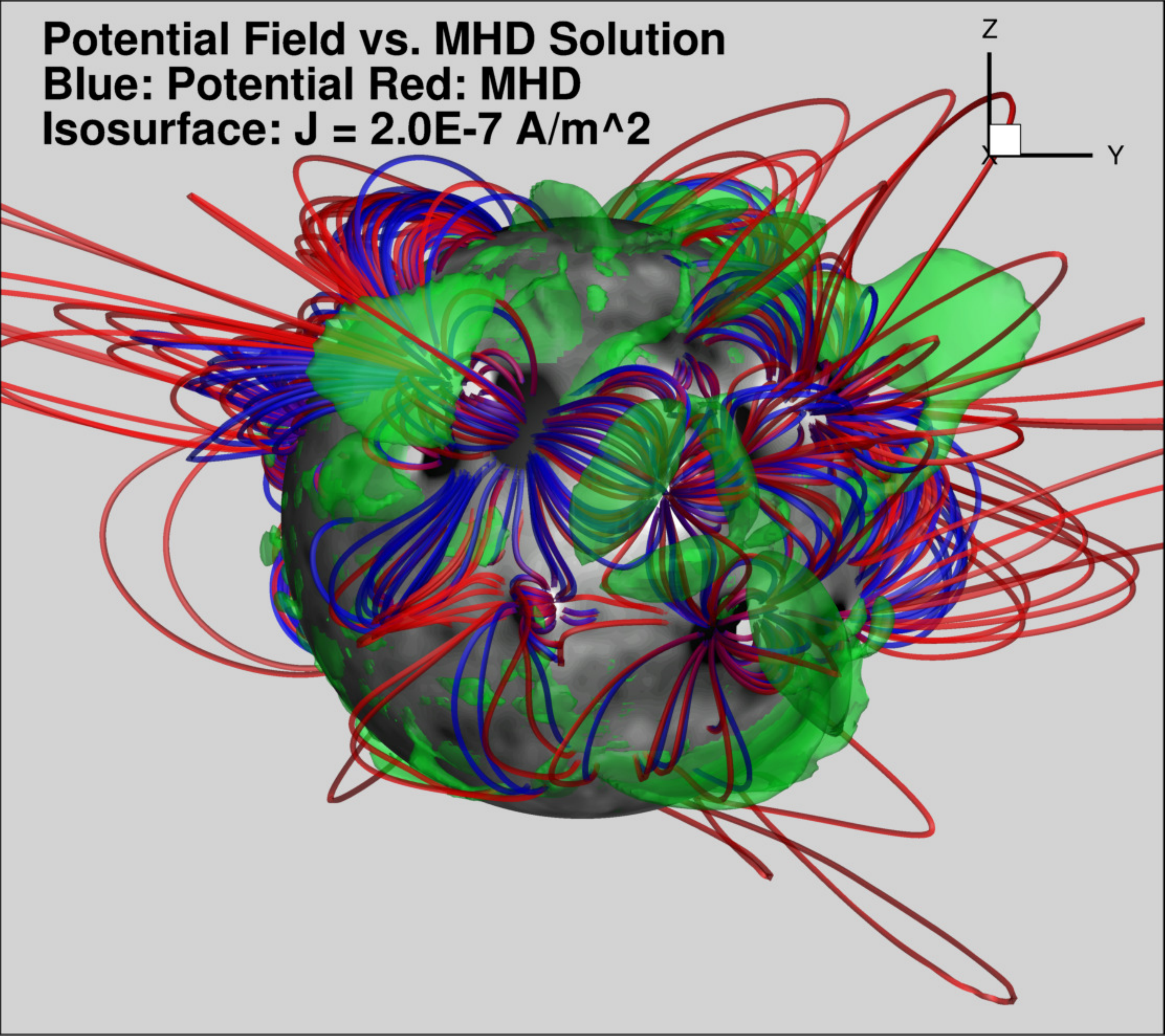}
\includegraphics[scale=0.21]{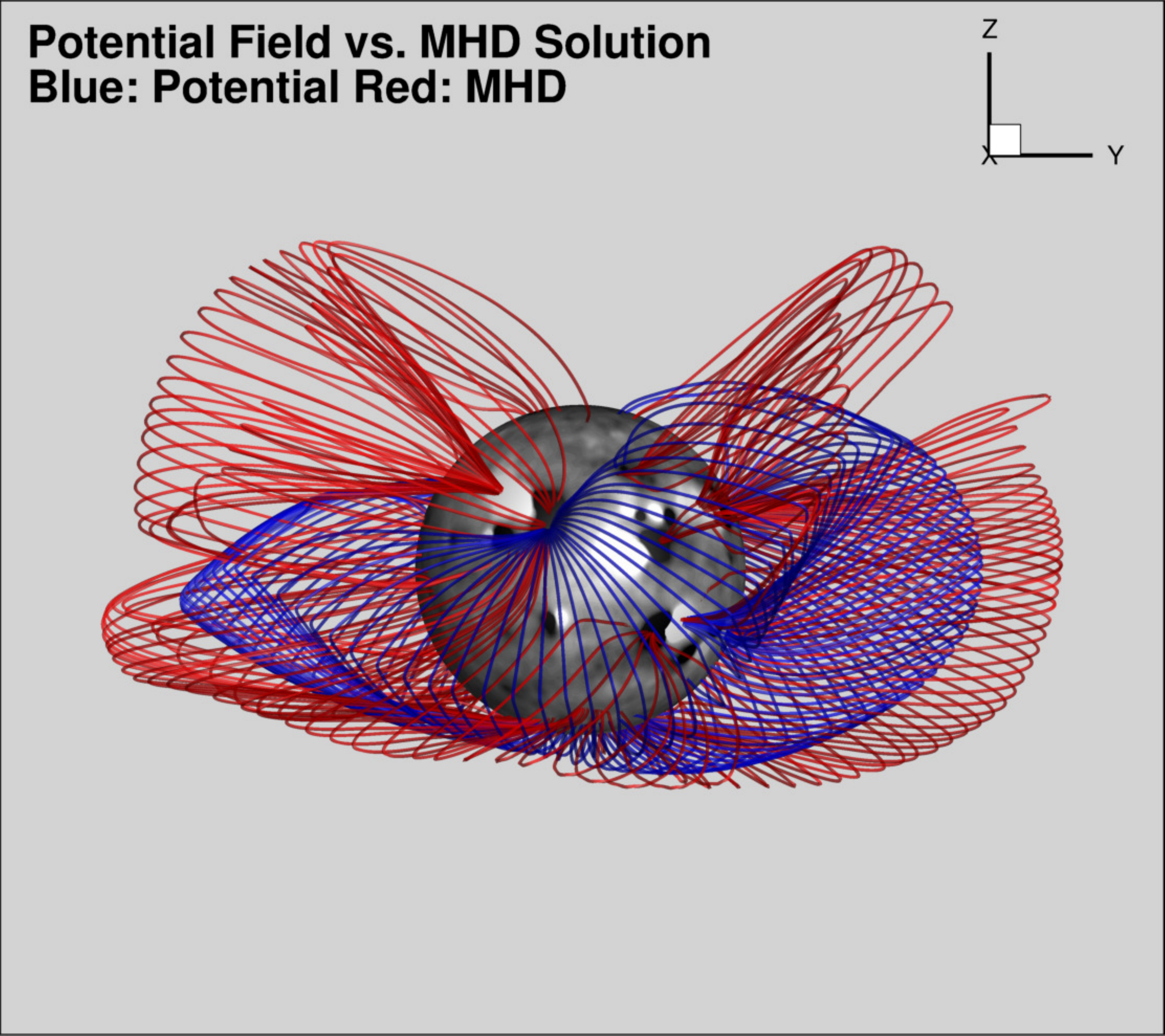}
\end{array}$
\end{center}
\caption{The comparison between the initial PFSS and final steady-state MHD solution of the 3D field configuration for near-Sun closed field (left panel) and large-scale helmet streamer belt (right panel). The blue field lines represent the potential field solution, and the red field lines represent the MHD solution. The green isosurface represents the current density $|{\bf J}|$ = 2$\times$10$^{-7}$ A m$^{-2}$. The helmet streamer field lines are selected according to the HCS locations at 2.5 R$_{\odot}$ for PFSS solution and 3.0 R$_{\odot}$ for MHD solution.}
\end{figure}

\newpage
\begin{figure}[h]
\begin{center}$
\begin{array}{c}
\includegraphics[scale=0.4]{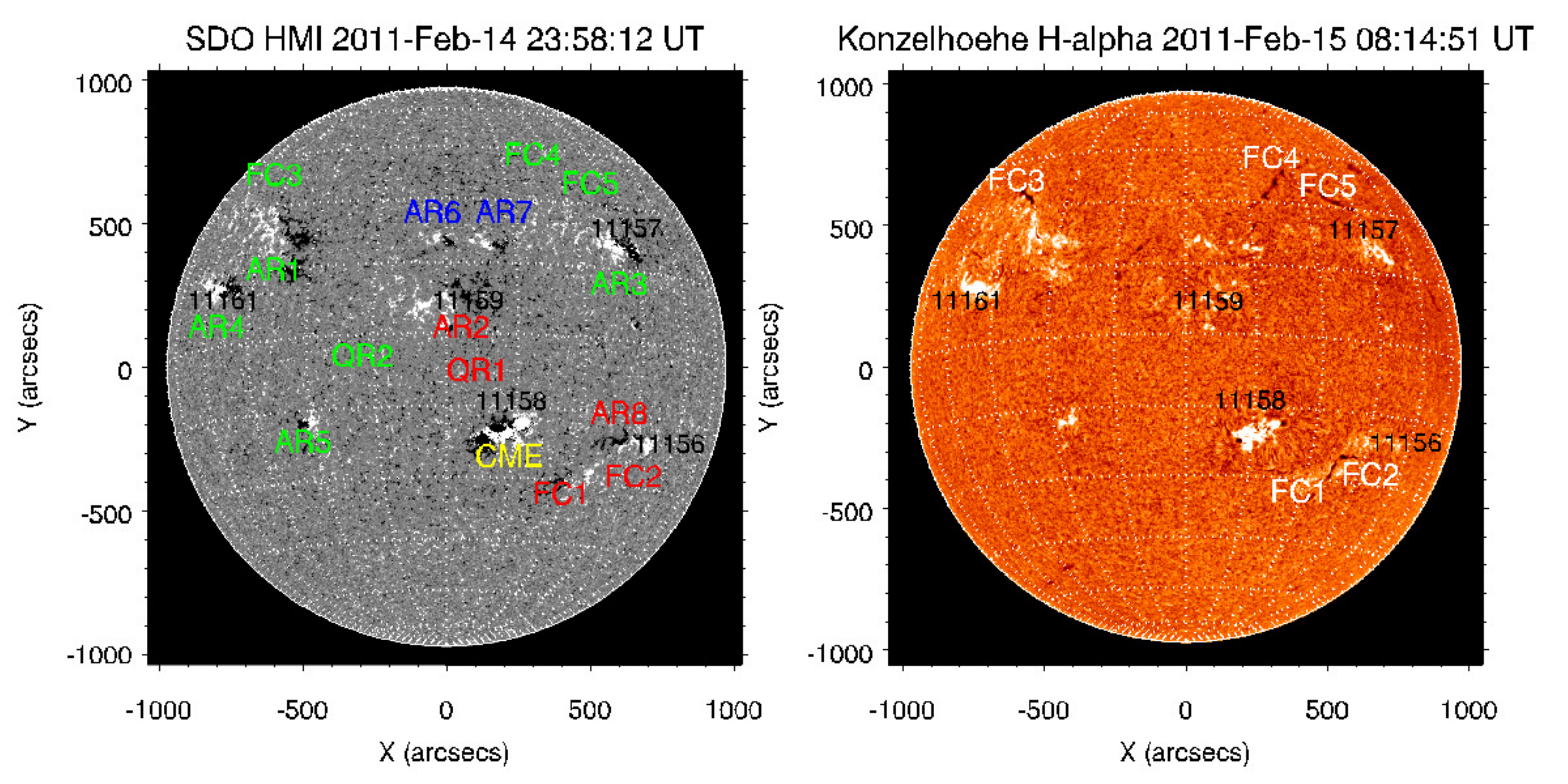}
\end{array}$
\end{center}
\caption{Left panel: solar structures overlaid on the HMI magnetogram of 2011 February 14 23:58:12 UT. Right panel: H$_{\alpha}$ observations on 2011 February 15 08:14:51 UT showing the positions of filament structures. The colors of the structures represent Type I (red), Type II (green), and Type III (blue) impacts described in \S 4.2.}
\end{figure}

\newpage
\begin{figure}[h]
\begin{center}$
\begin{array}{c}
\includegraphics[scale=0.65]{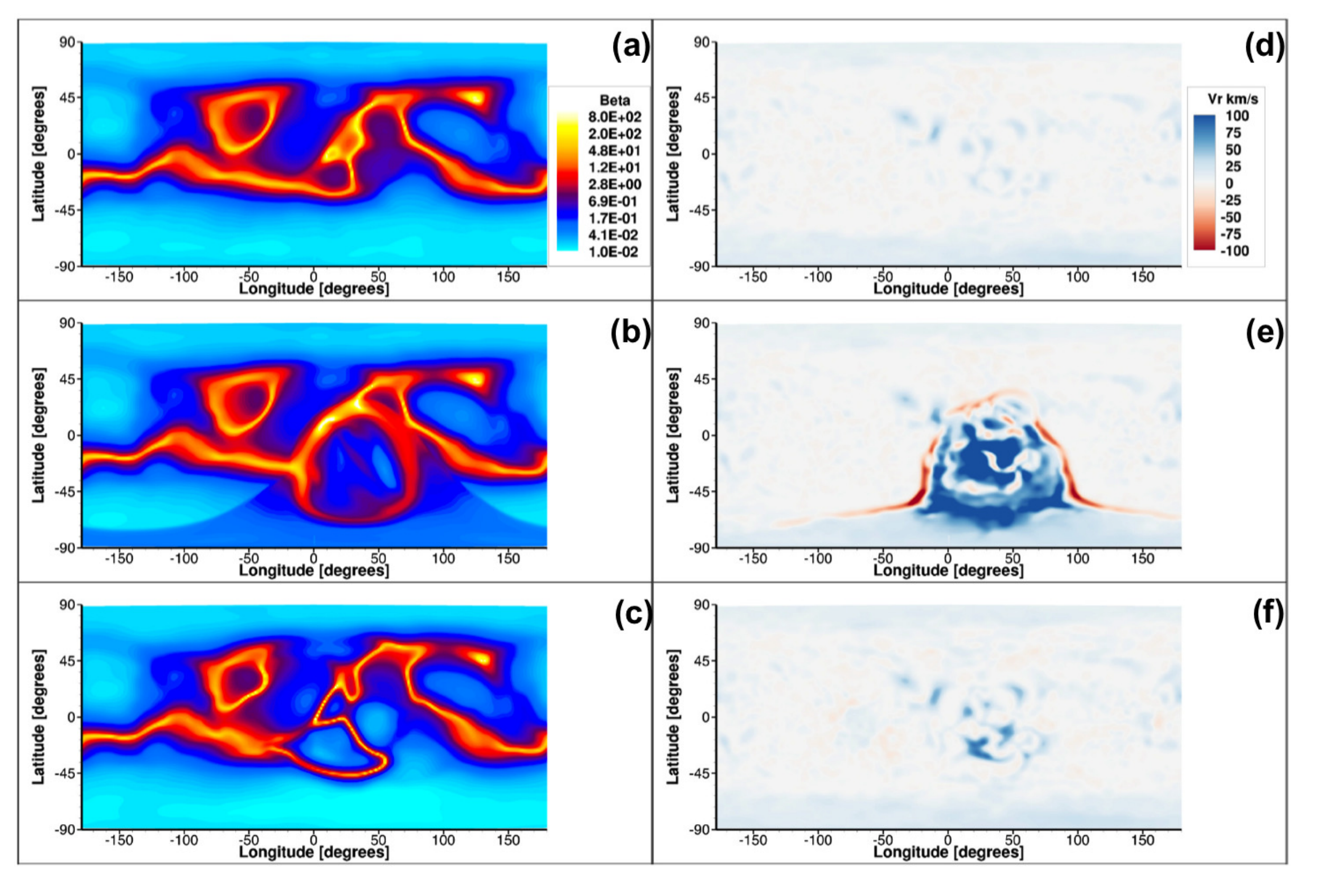}
\end{array}$
\end{center}
\caption{Panels (a)-(c): plasma-$\beta$ at 2.5 R$_{\odot}$ for t = 0, 10, and 60 min after the eruption. Panels (d)-(f): radial velocity at 42 Mm for t = 0, 10, and 60 min after the eruption. The red and blue colors represent downward and upward flows, respectively. Simulation data from Run 1 is used.}
\end{figure}

\newpage
\begin{figure}[htb]
\begin{center}$
\begin{array}{cc}
\vspace{-7pt}
\includegraphics[scale=0.25]{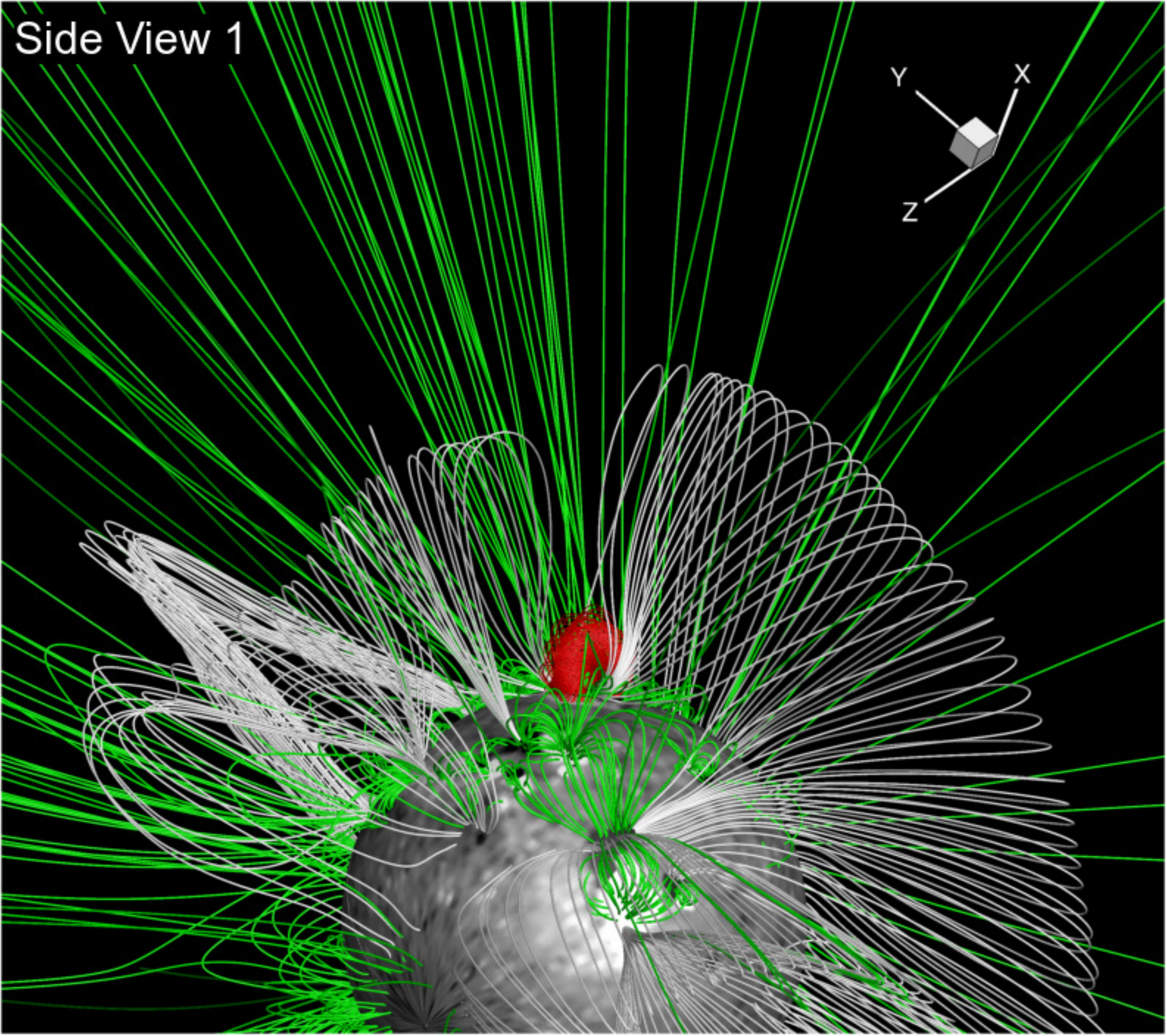}
\includegraphics[scale=0.25]{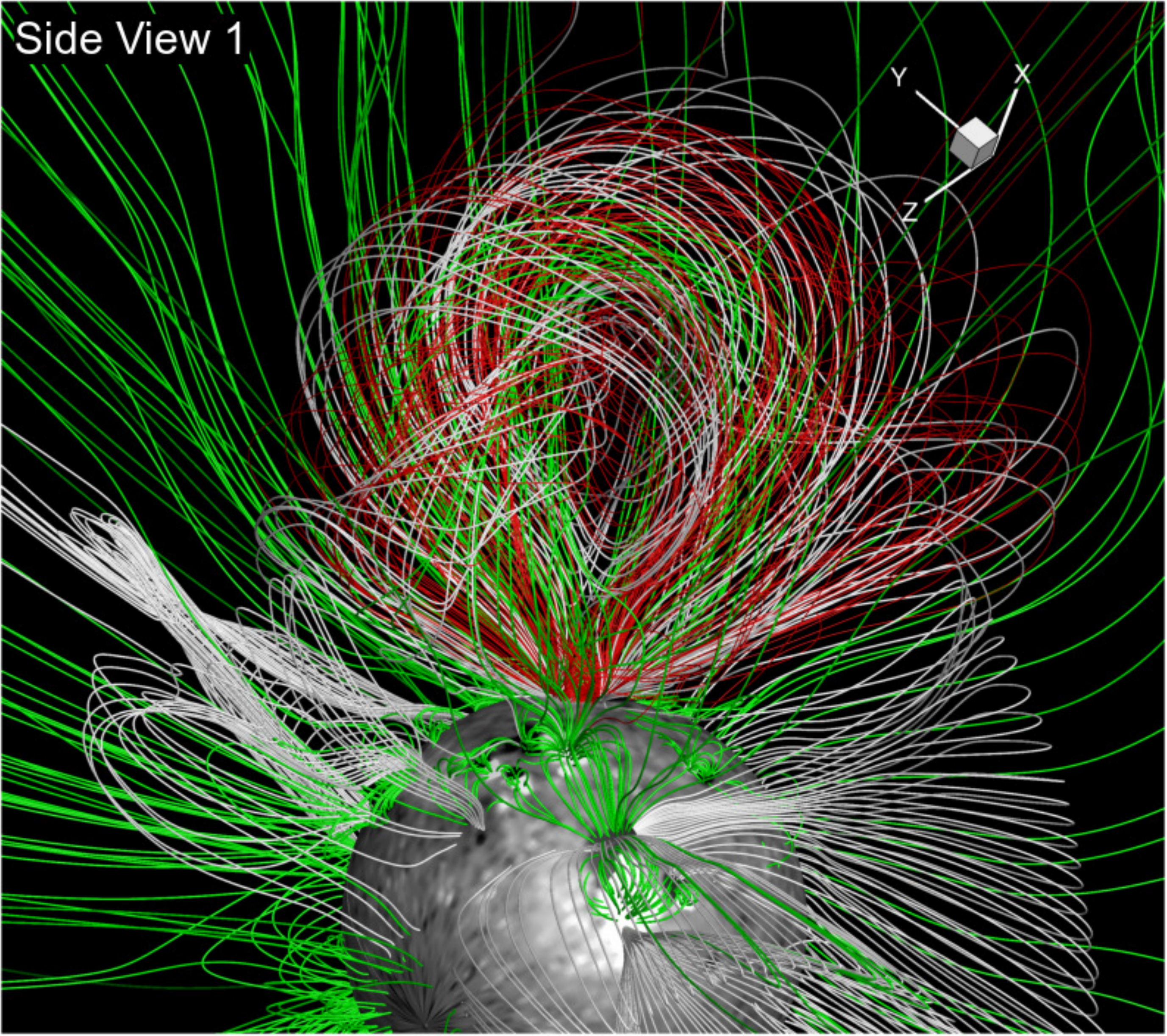}\\
\vspace{-7pt}
\includegraphics[scale=0.25]{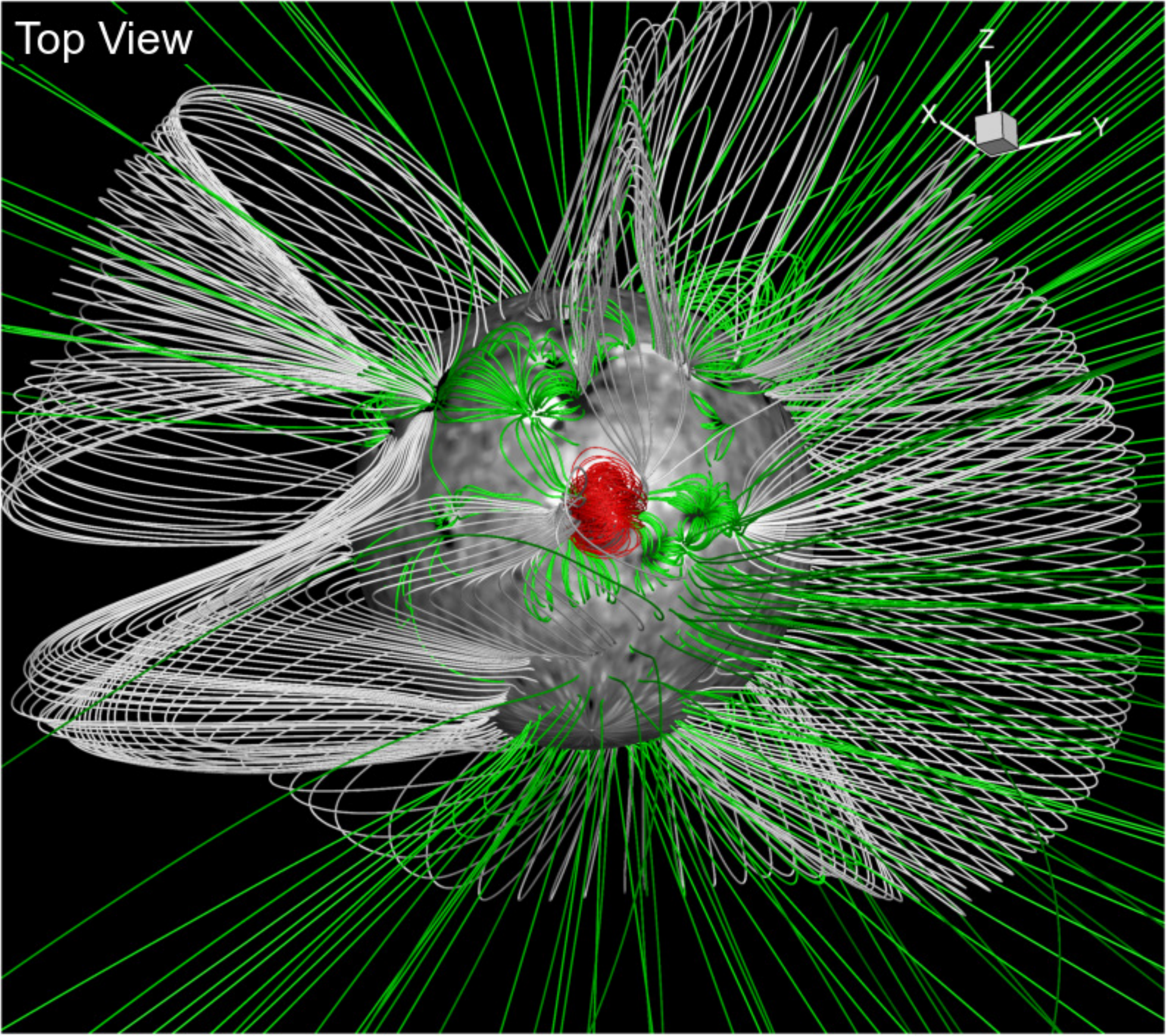}
\includegraphics[scale=0.25]{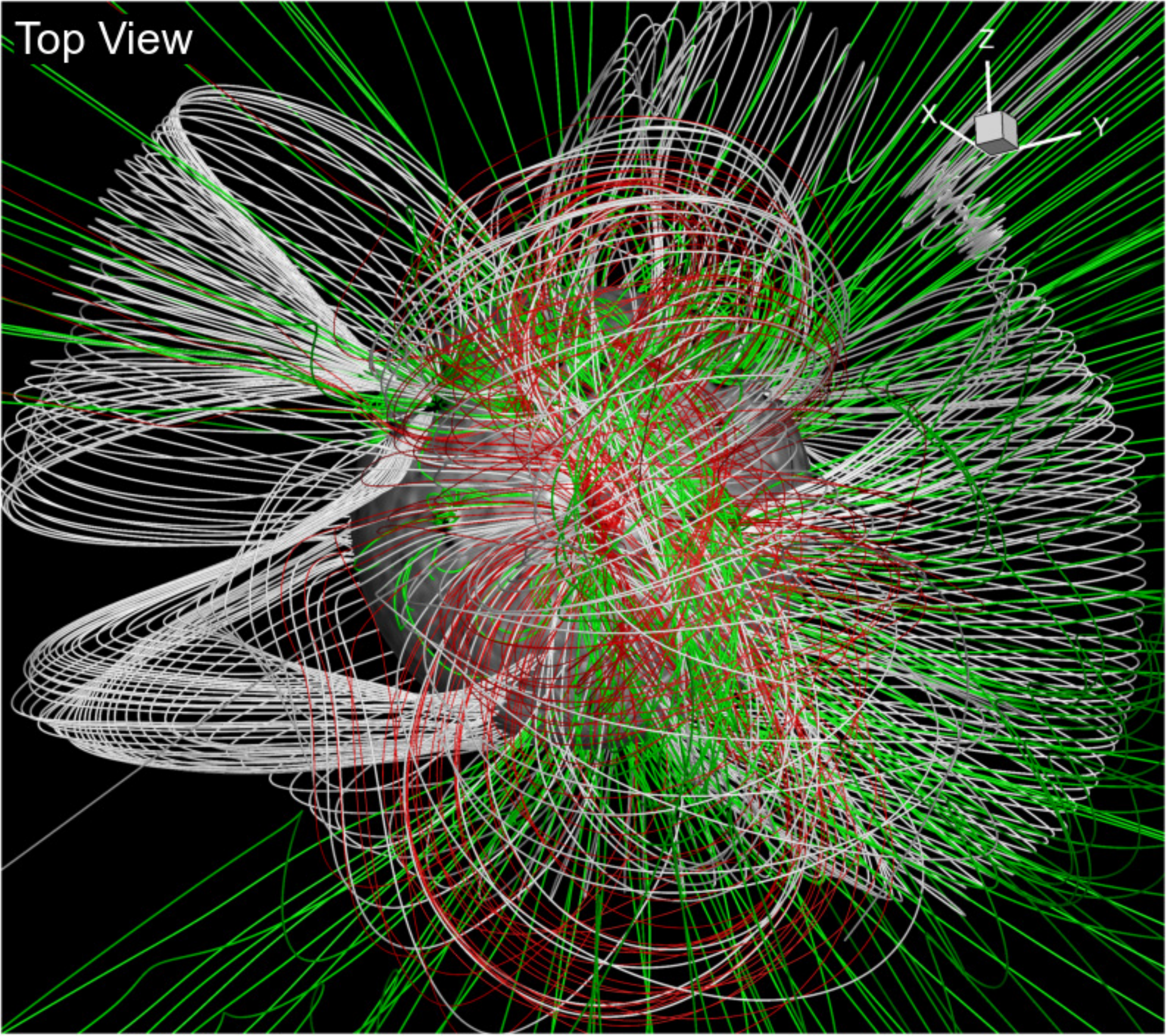}\\
\includegraphics[scale=0.25]{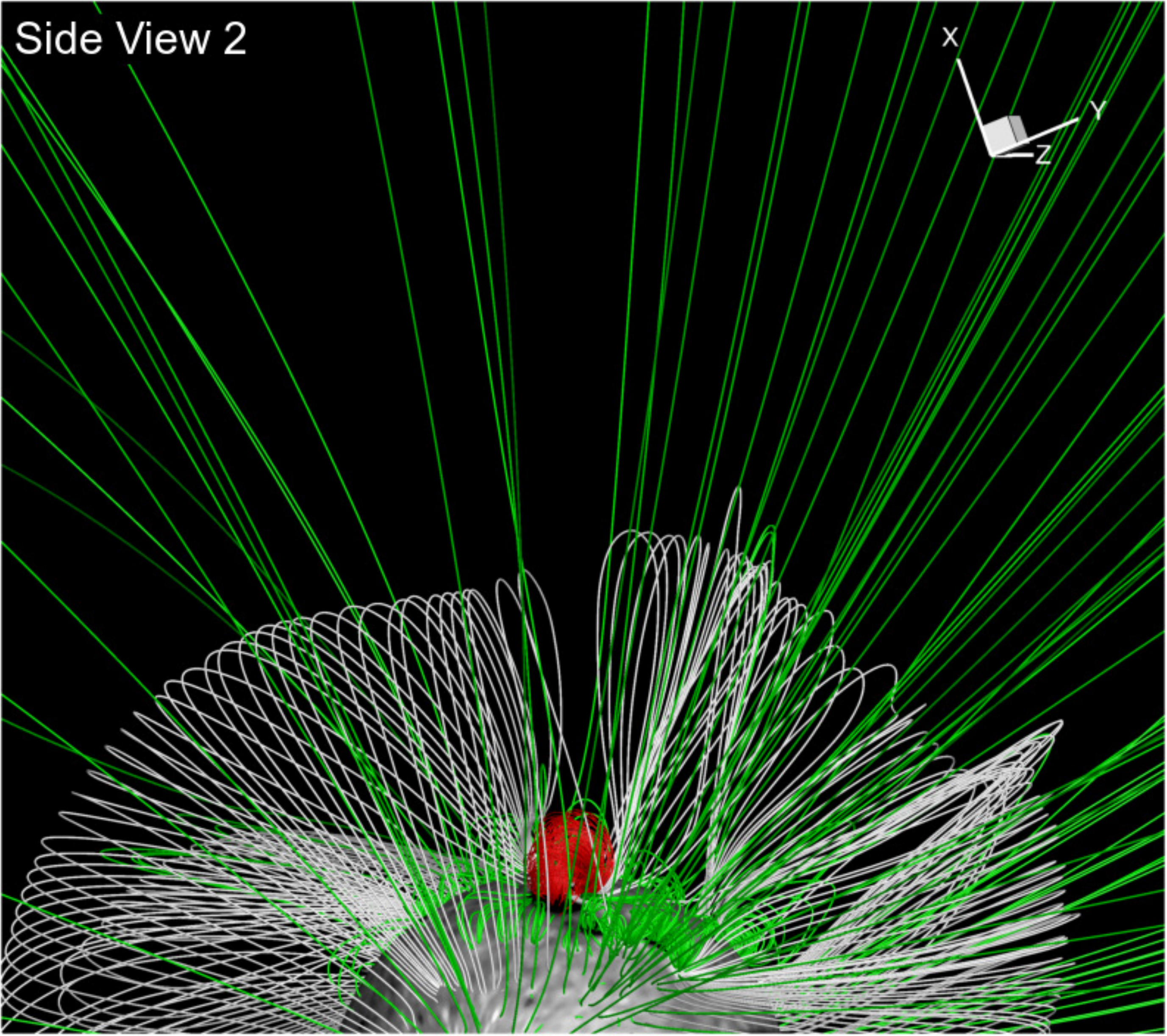}
\includegraphics[scale=0.25]{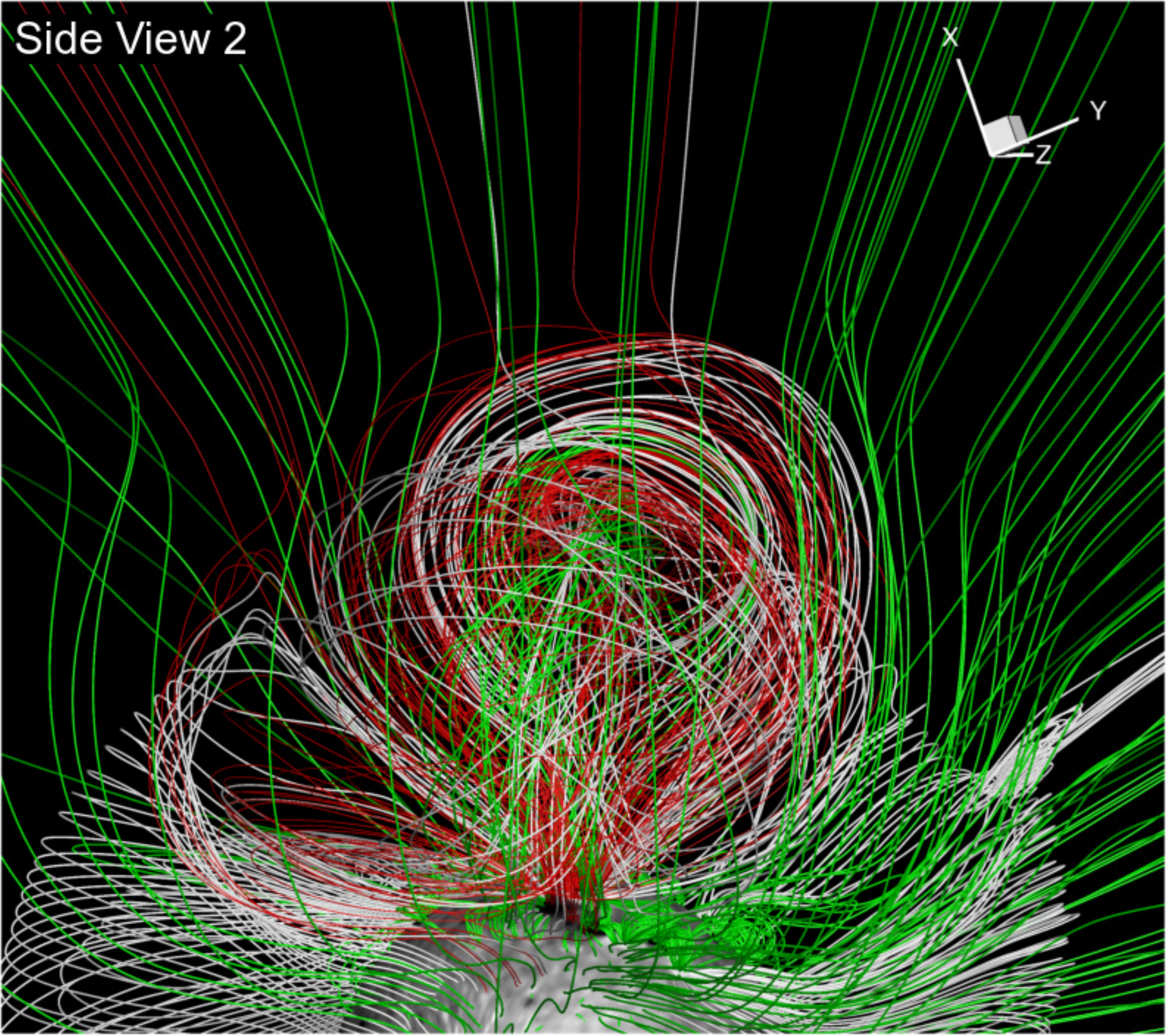}
\end{array}$
\end{center}
\caption{Magnetic field configuration from three different point of views at t = 0 (left column) and t = 15 min (right column). Run 1 data is used. The red, white, and green field lines represent flux rope field lines, large-scale helmet streamers, and field lines from surrounding active regions and open field. The coloring of field lines at t = 15 min is determined by the initial foot point location of individual field lines.  (An animation of this figure is available online)}
\end{figure}

\newpage
\begin{figure}[h]
\begin{center}$
\begin{array}{c}
\includegraphics[scale=0.32]{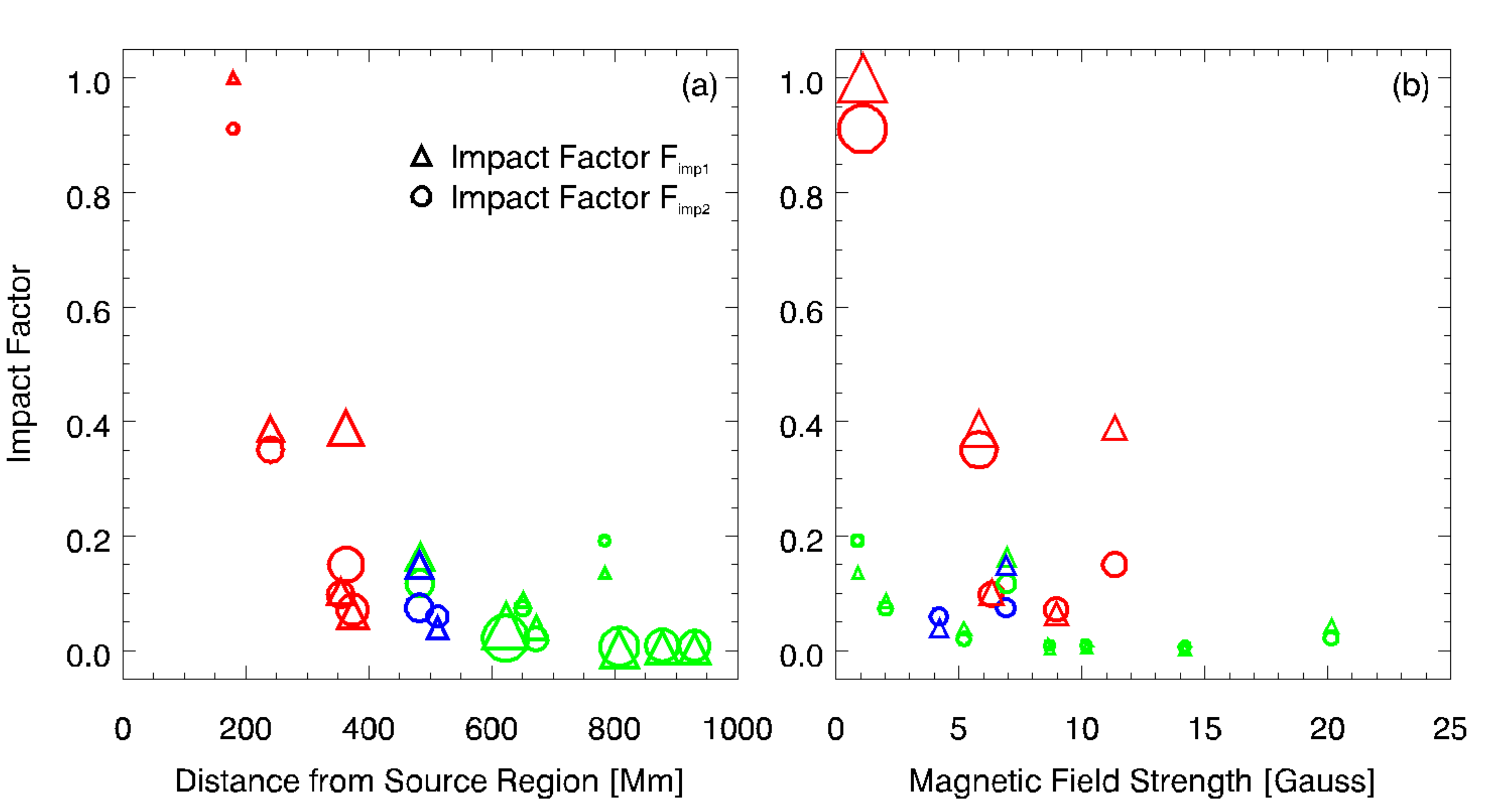}
\end{array}$
\end{center}
\caption{Impact factors $F_{{\rm imp1}}$ and $F_{{\rm imp2}}$ [Eq.(5) and Eq.(6)] as a function of (a) the distance from the source region, and (b) magnetic field strength of the structures. The colors represent Type I (red), Type II (green), and Type III (blue) impacts described in \S 4.2. In panel (a), larger symbol sizes represent stronger magnetic field strengths, and in panel (b), the symbol size indicates distance from the source region, with larger symbols corresponding to solar structures that are closer.}
\end{figure}

\newpage
\begin{figure}[h]
\begin{center}$
\begin{array}{c}
\includegraphics[scale=0.6]{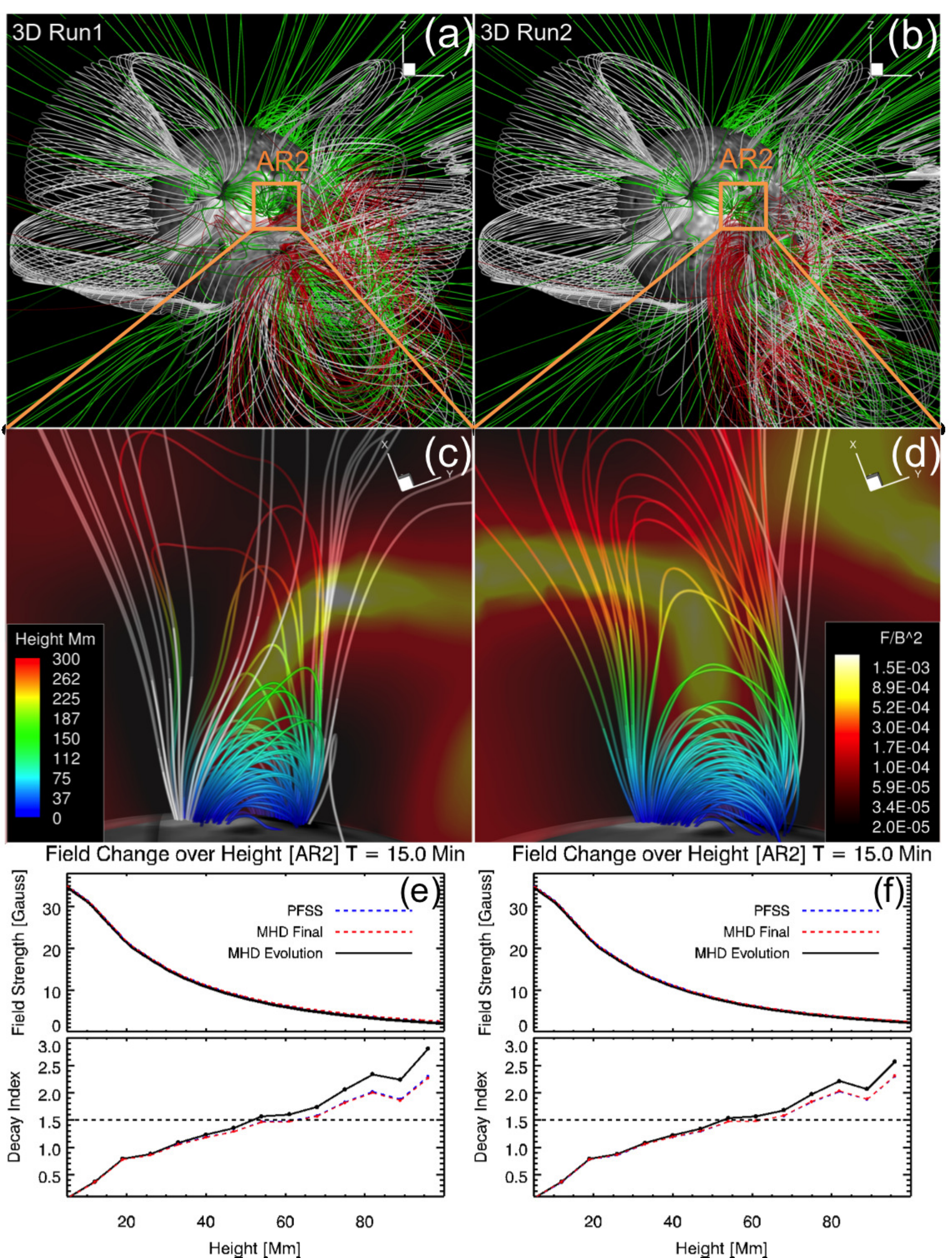}
\end{array}$
\end{center}
\caption{Panels (a) and (b) show the global magnetic field configuration of AR2 at t = 15 min for Runs 1 and 2. The red, white, and green field lines represent flux rope field lines, large-scale helmet streamers, and field lines from surrounding active regions and open field, respectively. Panels (c) and (d) show an enlarged view of AR2 as viewed from the side. The background shows the normalized Lorentz force and the coloring of the field lines shows the height. The white field lines represent the field lines reconnected with the erupting flux rope. Panels (e) and (f) show plots of the magnetic field strength and decay index along the height between 5 and 100 Mm (black solid lines). The PFSS and MHD final solutions are shown with blue and red dashed lines (which overlap in this case), respectively. (An animation of this figure is available online)}
\end{figure}

\newpage
\begin{figure}[h]
\begin{center}$
\begin{array}{c}
\includegraphics[scale=0.65]{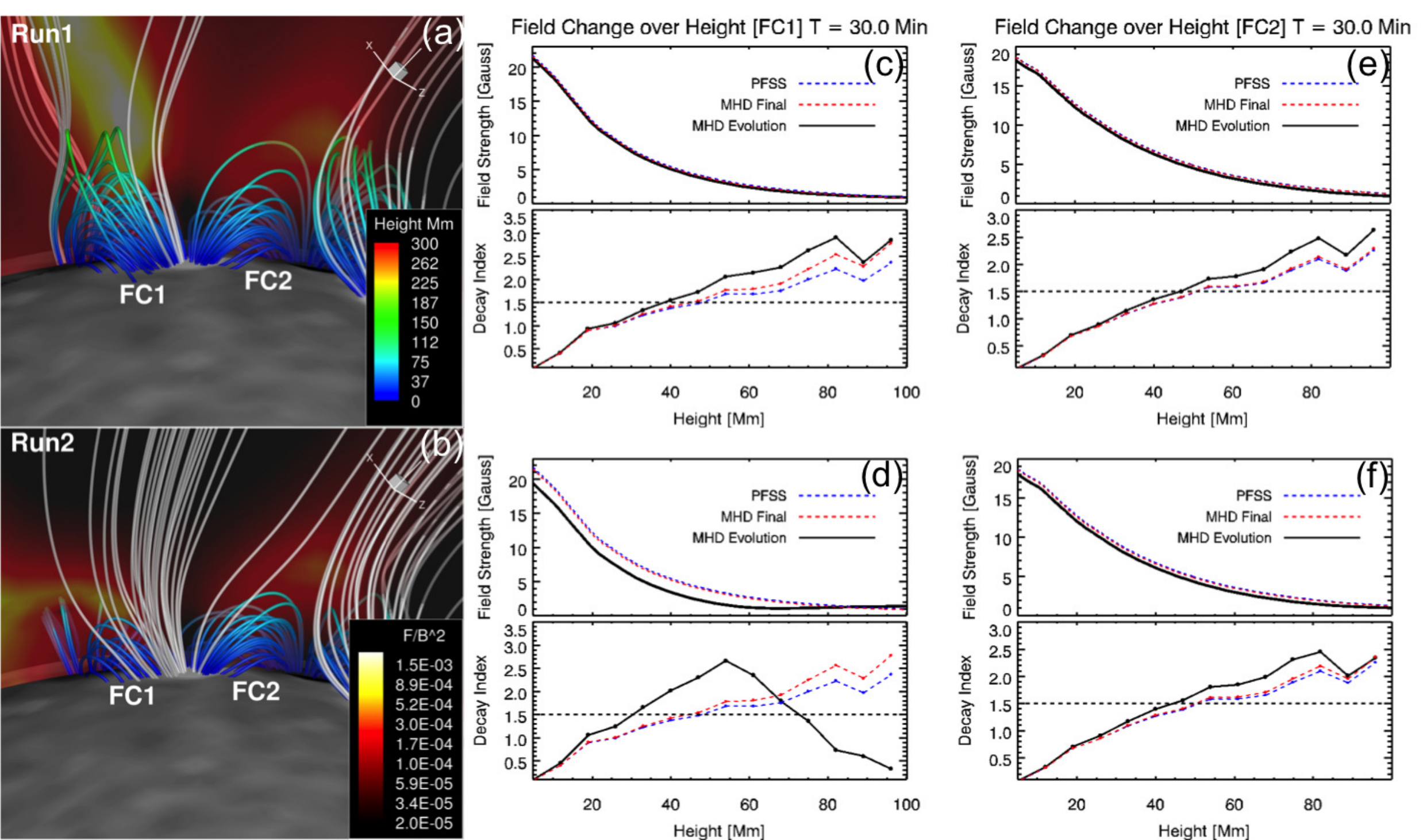}
\end{array}$
\end{center}
\caption{The magnetic field configuration and decay index of FC1 and FC2 at t = 30 min under different flux rope orientations (Run 1 and 2). Panels (a) and (b) show the zoom-in view for FC1 and FC2. The background shows the normalized Lorentz force and the coloring of the field lines shows the height. The white field lines represent the field lines reconnected with the erupting flux rope. Panels (c) through (f) show the magnetic field strength and decay index along the height between 5 and 100 Mm (black solid lines). The PFSS and MHD final solutions are shown with blue and red dashed lines, respectively. (An animation of this figure is available online)}
\end{figure}

\newpage
\begin{figure}[h]
\begin{center}$
\begin{array}{cc}
\includegraphics[scale=0.21]{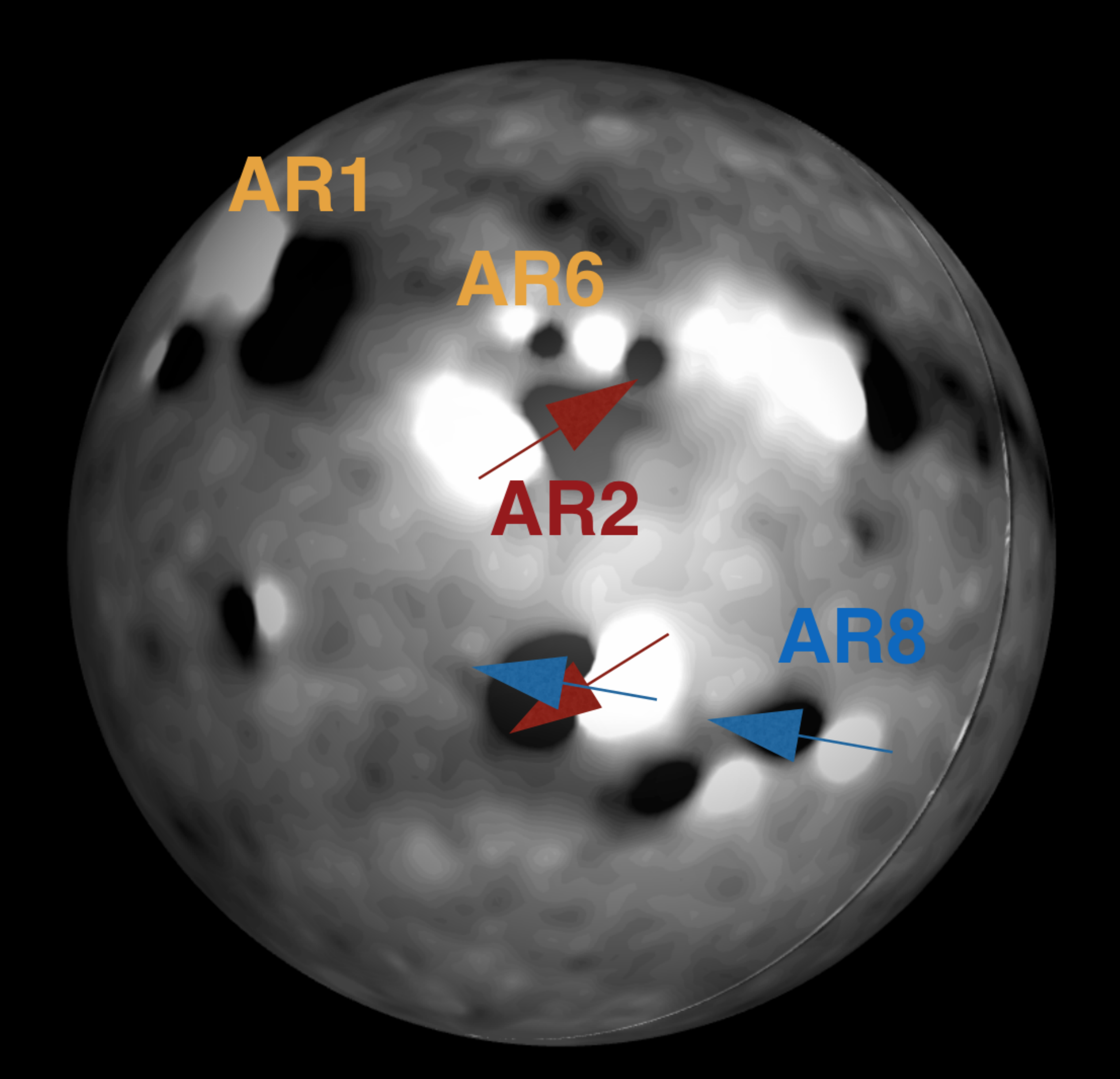}
\includegraphics[scale=0.19]{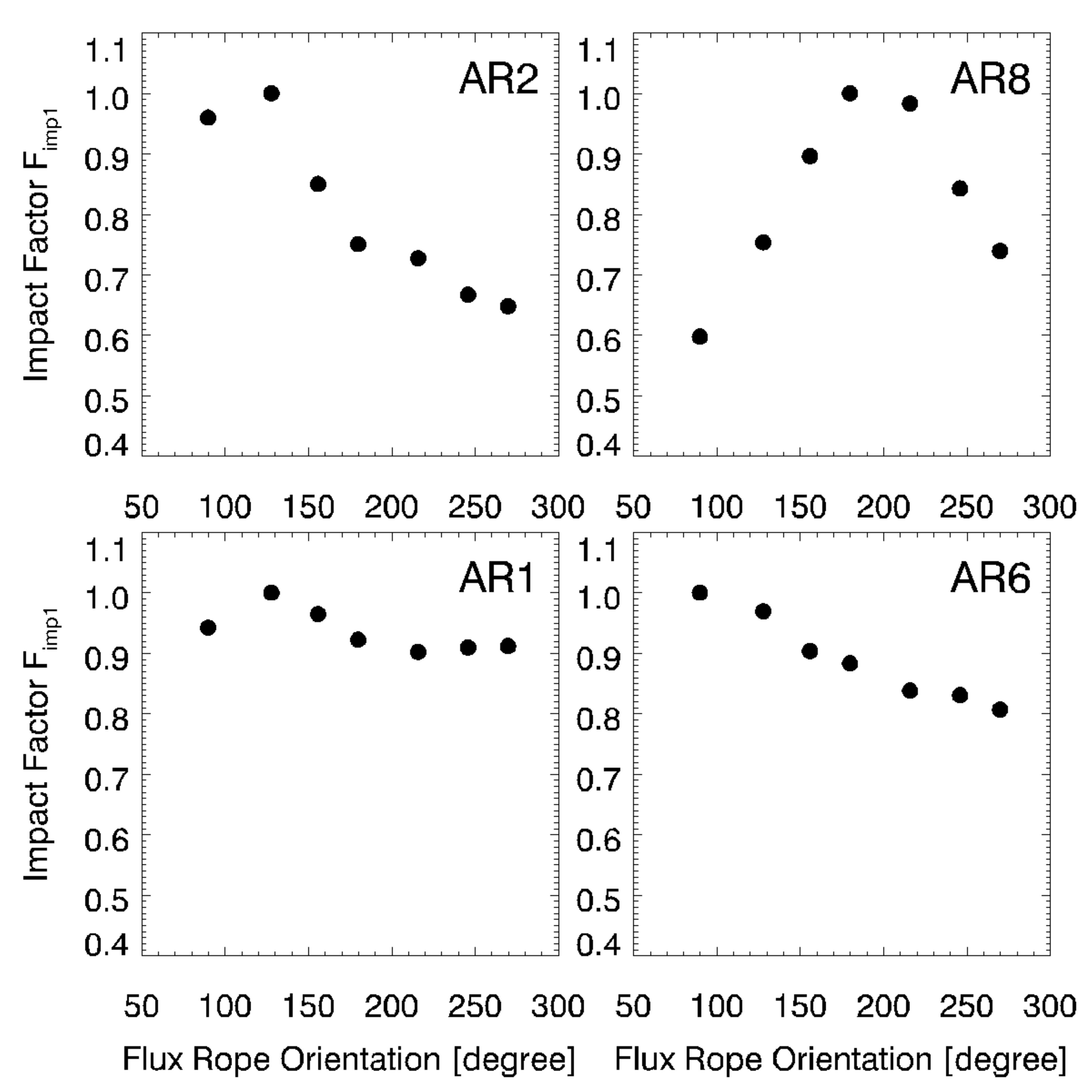}
\end{array}$
\end{center}
\caption{The image shows the location of active regions as well as the largest impact configuration for AR2 (red) and AR8 (blue). The plots show the orientation dependence of CME impact on AR1, AR2, AR6, and AR8. The impact factor I is used and scaled to the maximum impact factor of each structure.}
\end{figure}

\newpage
\begin{figure}[h]
\begin{center}$
\begin{array}{cc}
\includegraphics[scale=0.65]{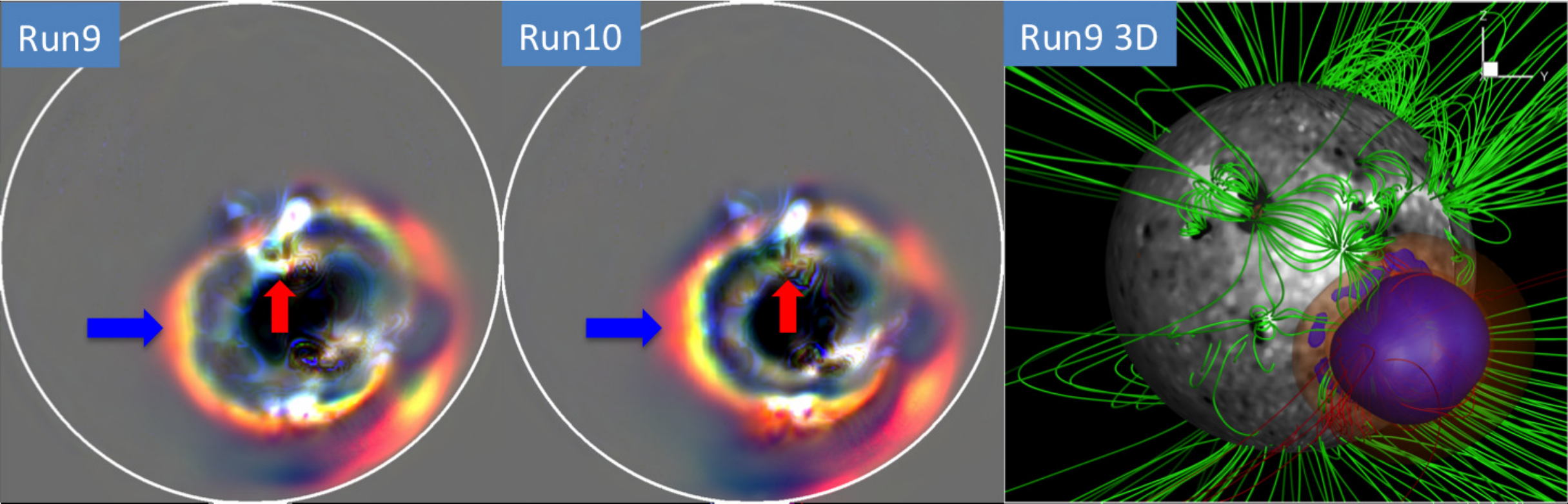}
\end{array}$
\end{center}
\caption{The simulated EUV waves and the corresponding 3D field configuration at t = 6 min. The synthesized images are produced by tri-ratio method with reference image at t = 4 min. The tri-color channels are AIA 211~\AA\ (red), AIA 193~\AA\ (green), and AIA171~\AA\ (blue). The ratio in each channel is scaled to 1$\pm$0.3. The blue arrow marks the location of fast-mode wave, and the red arrow marks the location where the flux rope field reconnects with AR2 field. The red and green field lines in the right panel represent flux rope and surrounding active region/open field, respectively. The red isosurface represents density ratio of 1.05 between t = 6 min and t = 4 min solution. The blue isosurface represents a density ratio of 0.8.}
\end{figure}

\newpage
\begin{figure}[h]
\begin{center}$
\begin{array}{c}
\includegraphics[scale=0.65]{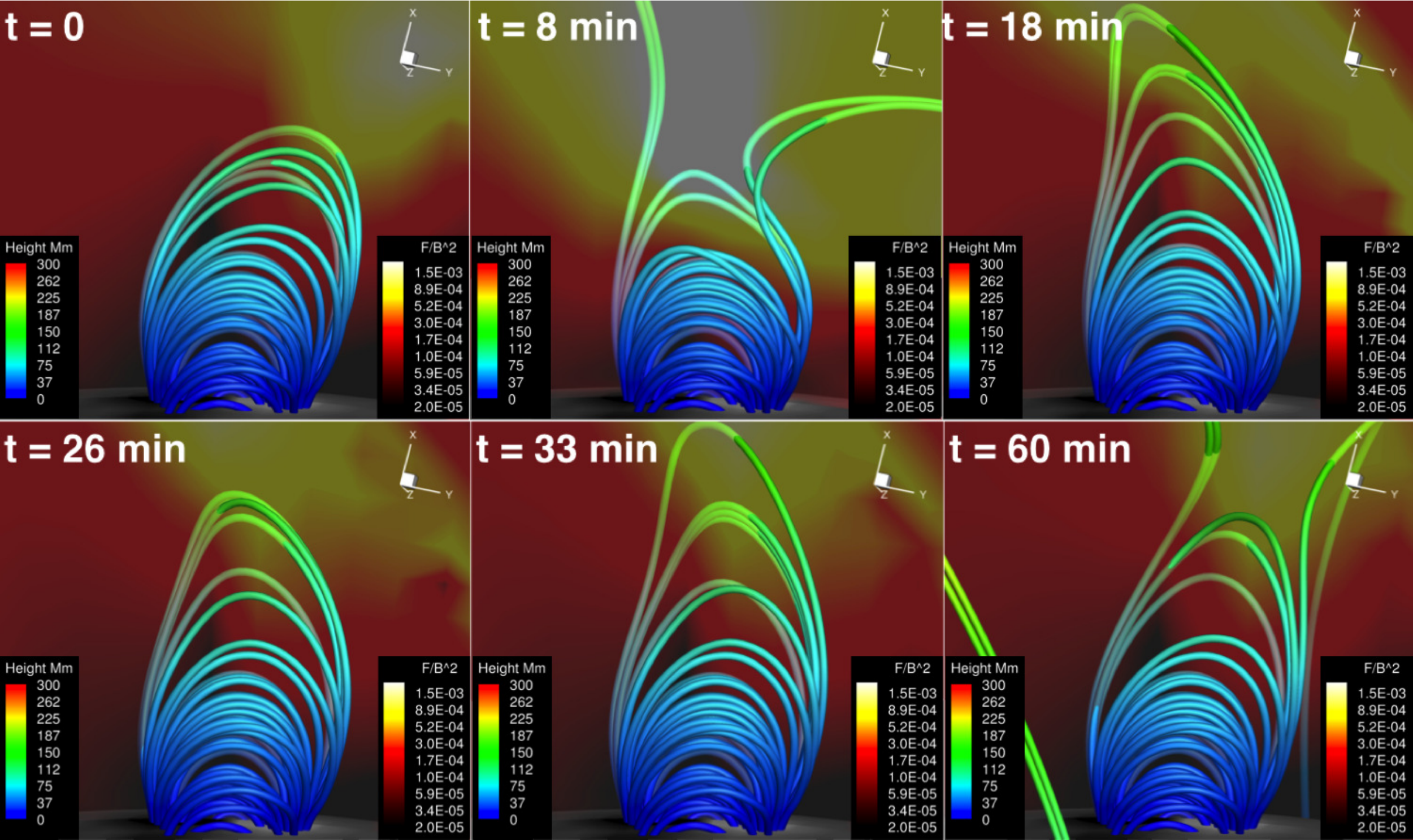}
\end{array}$
\end{center}
\caption{Recovery phase evolution of AR5. The background shows the normalized Lorentz force and the coloring of the field lines shows the height in Mm. The eruption occurs at t = 0.}
\end{figure}

\newpage
\begin{figure}[h]
\begin{center}$
\begin{array}{c}
\includegraphics[scale=0.3]{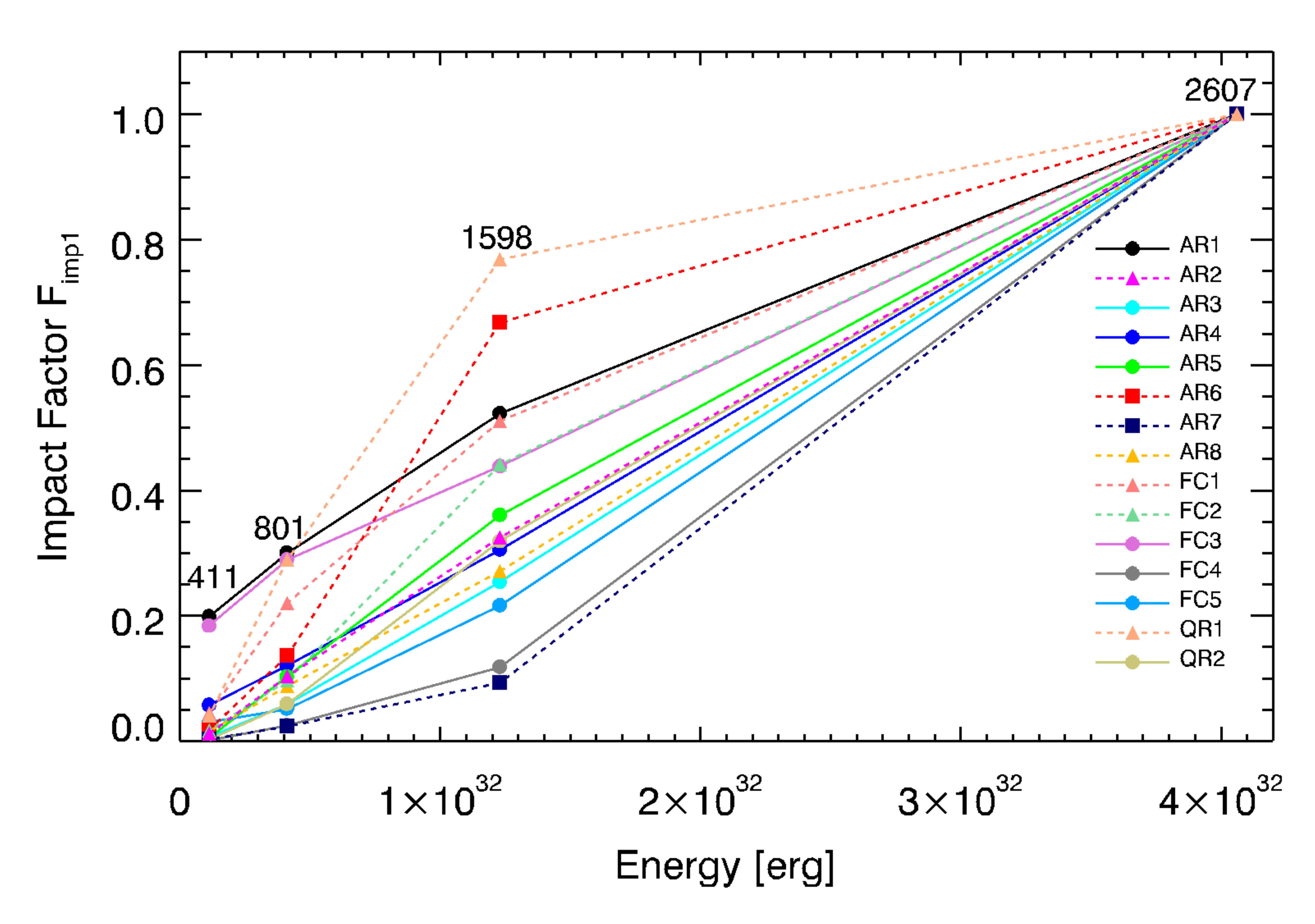}
\end{array}$
\end{center}
\caption{Energy dependence of CME impacts on different structures. The impact factor $F_{{\rm imp1}}$ is used and normalized to the maximum impact factor of each structure. Type I, II, and III impacts are represented by triangle, circle, and square symbols, respectively. Also, Type I/III and Type II impacts can be distinguished by dashed and solid lines. The numbers above each energy show the final CME speed in the unit of km s$^{-1}$.}
\end{figure}

\newpage
\begin{figure}[h]
\begin{center}$
\begin{array}{c}
\includegraphics[scale=0.63]{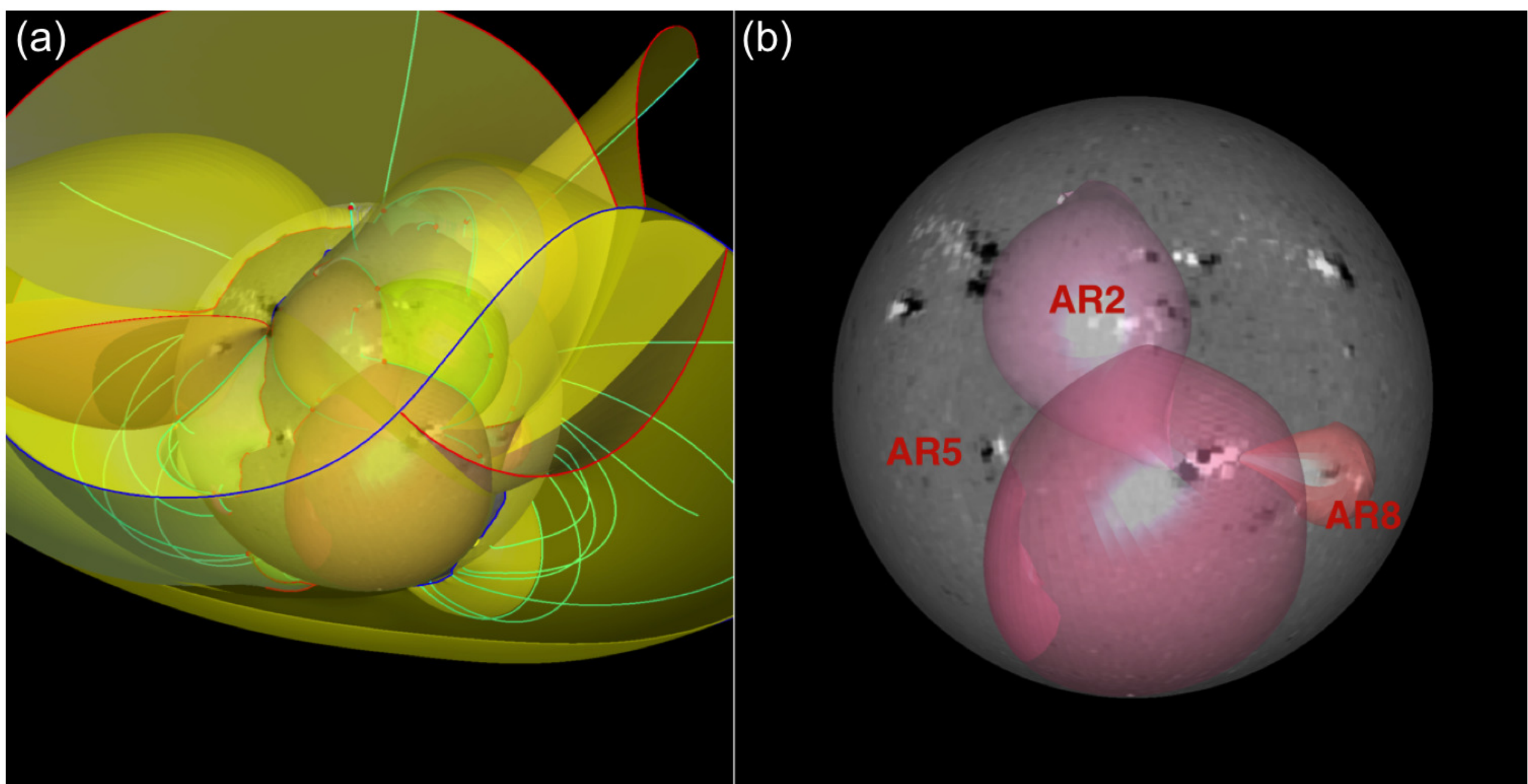}
\end{array}$
\end{center}
\caption{(a) Topological structures calculated from a PFSS model. The yellow surface is the PFSS helmet surface dividing domains of open and closed flux. The colored surfaces underneath are  topological domains defined by null points (red dots) and their spine field lines (light blue). The blue line shows the null line on the source surface. The red lines are the intersections of vertical separatrix surfaces with the solar surface and source surface. (b) The selection of topological domains from the left panel that directly connect to the CME source region.}
\end{figure}

\end{document}